\documentclass{JFM-FLM_Au}

\usepackage{caption}
\usepackage{subcaption}
\usepackage{graphicx}
\usepackage[absolute,overlay]{textpos}
\usepackage[percent]{overpic}
\usepackage[normalem]{ulem}

\lefttitle{Z. Long, P. Gao}
\righttitle{Journal of Fluid Mechanics}

\title{Migration and spreading of a droplet driven by a chemical step}

\author{Zhuo Long\aff{1}, Peng Gao\aff{1}}

\affiliation{\aff{1}Department of Modern Mechanics, University of Science and Technology of China, Hefei, Anhui 230026, PR China}

\corresau{Peng Gao, \email{gaopeng@ustc.edu.cn}}

\begin{document}
\maketitle

\begin{abstract}
The chemical step is an elementary pattern in chemically heterogeneous substrates, featuring two regions of different wettability separated by a sharp border. Within the framework of lubrication theory, we investigate droplet motion and the contact-line dynamics driven by a chemical step, with the contact-line singularity addressed by the Navier slip condition. For both two-dimensional (2D) and three-dimensional (3D) droplets, two successive stages are identified: the migration stage, when the droplet traverses both regions, and the asymmetric spreading stage, when the droplet spreads on the hydrophilic region while being constrained by the border. For 2D droplets, we present a matched asymptotic analysis which agrees with numerical solutions. In the migration stage, a 2D droplet can exhibit translational motion with a constant speed. In the asymmetric spreading stage, the contact line at the droplet rear is pinned at the border. We show that a boundary layer still exists near the pinned contact line, across which the slope is approximately constant, whereas the curvature would diverge in the absence of slip. For 3D droplets, our numerical simulations show that the evolution is qualitatively analogous to the 2D case, while being significantly affected by the lateral flow. The droplet length and width exhibit non-monotonic variations due to the lateral flow. Eventually, the droplet detaches from the border and approaches equilibrium at the hydrophilic substrate. Additionally, we demonstrate that the variation of the apparent contact angle at the instant of border contact only affects the early stage of droplet migration.
\end{abstract}




\section{Introduction}  \label{Sec:Intro}

The motion of a sessile droplet on a solid substrate is a ubiquitous phenomenon and an intriguing challenge of scientific researches \citep{degennes1985, bonn2009, snoeijer2013}. It is also of great importance in industrial processes such as inkjet printing \citep{lohse2022} and lab-on-a-chip technologies \citep{stone2004}.  Sessile droplet motion can be driven by external effects such as light \citep{ichimura2000}, electric fields \citep{hartmann2022}, and ultrasound \citep{liu2025}, or designed substrate gradients including wettability, curvature \citep{lv2014}, and temperature \citep{sui2014}. For reviews, refer to \citet{malinowski2020}, \citet{tenjimbayashi2022} and \citet{moragues2023}.

Most real-world solid surfaces are chemically heterogeneous, and the wettability distribution on such surfaces provides guidance for the sessile droplets. A notable example is the bioinspired wettability gradient surfaces in water collection. \citet{parker2001} investigated the ability of desert beetles collecting water from fog-laden winds using the covering of their backs, and reported the pivotal role of alternating hydrophilic and hydrophobic regions. Inspired by such natural phenomena, \citet{bai2014} developed a water collecting surface with star-shaped wettability patterns, delivering significantly higher water-collection efficiency. 

In fundamental research, understanding how chemical heterogeneity influences fluid behaviour remains a central question. The term \textit{Chemically heterogeneous substrates} refers to surfaces that are macroscopically flat, yet exhibit spatially varying wettability due to microscopic material properties. Typical categories are \textit{wettability gradient surfaces}, where the wettability varies gradually \citep{greenspan1978, dossantos1995, thiele2004, ahmadlouydarab2014}, and \textit{wettability patterned surfaces}, characterized by sharp borders between hydrophilic and hydrophobic regions, forming patterns such as arrays \citep{cubaud2004} and stripes \citep{dulcey1991}. \textit{Stripe patterned surfaces} consist of alternating hydrophilic and hydrophobic long stripes. The effect of stripe width relative to the droplet size has been widely studied to examine anisotropic wetting \citep{jansen2012, jansen2014}, and to model rough surfaces \citep{iwamatsu2006, damle2017} or contact angle hysteresis \citep{wang2008, xu2011, xu2020}. In studies on droplet motion on stripe patterned surfaces, it has been shown that droplets can be directed along designed stripes \citep{zhao2001, chowdhury2019}. Driven by external forces such as gravity, droplets can cross wettability borders, where several types of depinning have been reported \citep{thiele2006, he2020}. Droplets sliding through wettability borders is a prototype problem for studying the stick-slip phenomenon, where the advancing contact line sticks when crossing the border from a hydrophilic stripe to a hydrophobic one, and slips otherwise \citep{wang2008, liu2017}, reducing the average droplet speed by an order of magnitude \citep{varagnolo2013, sbragaglia2014}. Recent research on dewetting has reported similar slip motion when contact line recedes from a hydrophilic region to a hydrophobic one \citep{wang2025a}. By examining droplet motion on a single wettability border, their results revealed the intrinsic distinction between dewetting and wetting. 

The \textit{chemical step} refers to an elementary pattern of heterogeneous substrates (particularly stripe patterned surfaces), characterized by two regions of different wettability separated by a sharp border. Despite its ubiquity, the behaviour of droplets on a chemical step remains poorly understood. Existing studies have mainly examined droplets on inclined substrates with a chemical step, with the hydrophobic region at the bottom. \citet{semprebon2016} conducted experiments on droplet sliding on a chemical step, and observed several droplet trajectories under different conditions including substrate inclination and border angle. Their results were coupled with contact angle hysteresis, since it is difficult to be eliminated in experiments. \citet{deconinck2017} developed an equilibrium theory to describe droplet morphology pinned at a wettability border. \citet{devic2019} derived criteria for pinning and depinning of small droplets, neglecting the normal gravity component. For larger droplets, their model showed that the normal force drives the droplet back towards the hydrophilic region. The dynamical process of a droplet sliding across a chemical step was not investigated until \citet{li2023} employed lattice Boltzmann simulations and investigated the transmission between kinetic and surface energy. Overall, previous studies have focused on the competing effects of the chemical step and gravity, yet droplet motion on a horizontal substrate with a chemical step has not been rigorously investigated.

A universal challenge in the study of droplet motion on substrates is the well-known Huh-Scriven paradox \citep{huh1971}, which can be addressed by moving contact line models \citep{snoeijer2013}. These models are typically characterized by a microscopic length scale, such as the slip length, rendering the wetting process inherently multiscale. The matched asymptotic analysis have proven particularly suitable for such problems, leading to several important results, including Tanner's law \citep{tanner1979} and the Cov--Voinov relation \citep{voinov1976, cox1986}. Furthermore, by introducing the Navier slip model, a matching procedure was developed by \citet{hocking1982} and \citet{lacey1982} in the study of droplet spreading. As for heterogeneous substrates, \citet{pismen2006} examined the moving two-dimensional (2D) droplet driven by a wettability gradient. \citet{vellingiri2011} employed a 2D matching procedure for general wettability distributions  characterized by smoothly varying, non-zero contact angles. Their model is extended to scenarios in three-dimensional (3D) \citep{savva2019} or with mass transfer \citep{groves2021, savva2021}. The matched asymptotic analysis is expected to be helpful in understanding the motion of droplets on a chemical step, while relevant investigation is still lacking. From the numerical perspective, the multiscale nature results to a huge computational cost \citep{sui2013, sui2014a}. A feasible method to reduce the cost is to impose a contact line model to resolve the microscale behaviours, and employ numerical simulations only in the macroscale evolutions \citep{qin2024}, which is adopted in the simulations of the present work. 

In the present study, we consider the fundamental process of droplet motion driven by a chemical step,  where a droplet on the less hydrophilic region initially contacts the wettability border and subsequently moves to the more hydrophilic region. The paper is organized as follows. \S\,\ref{Sec:Eqn} describes the 2D and 3D problem in the framework of lubrication theory. For the 2D problem discussed in \S\,\ref{Sec:2D}, two stages are distinguished and solved separately with a matching procedure, which is validated against numerical solutions. \S\,\ref{Sec:3D} performs numerical simulations for the 3D case and discusses the 3D effects. \S\,\ref{Sec:IC} presents the effects of the initial condition, where the droplet could be spreading before contacting with the wettability border. Some conclusions are drawn in \S~\ref{Sec:conclu}.

\section{Governing equations}\label{Sec:Eqn}

\begin{figure}
    \centering
    \includegraphics[width=.5\linewidth]{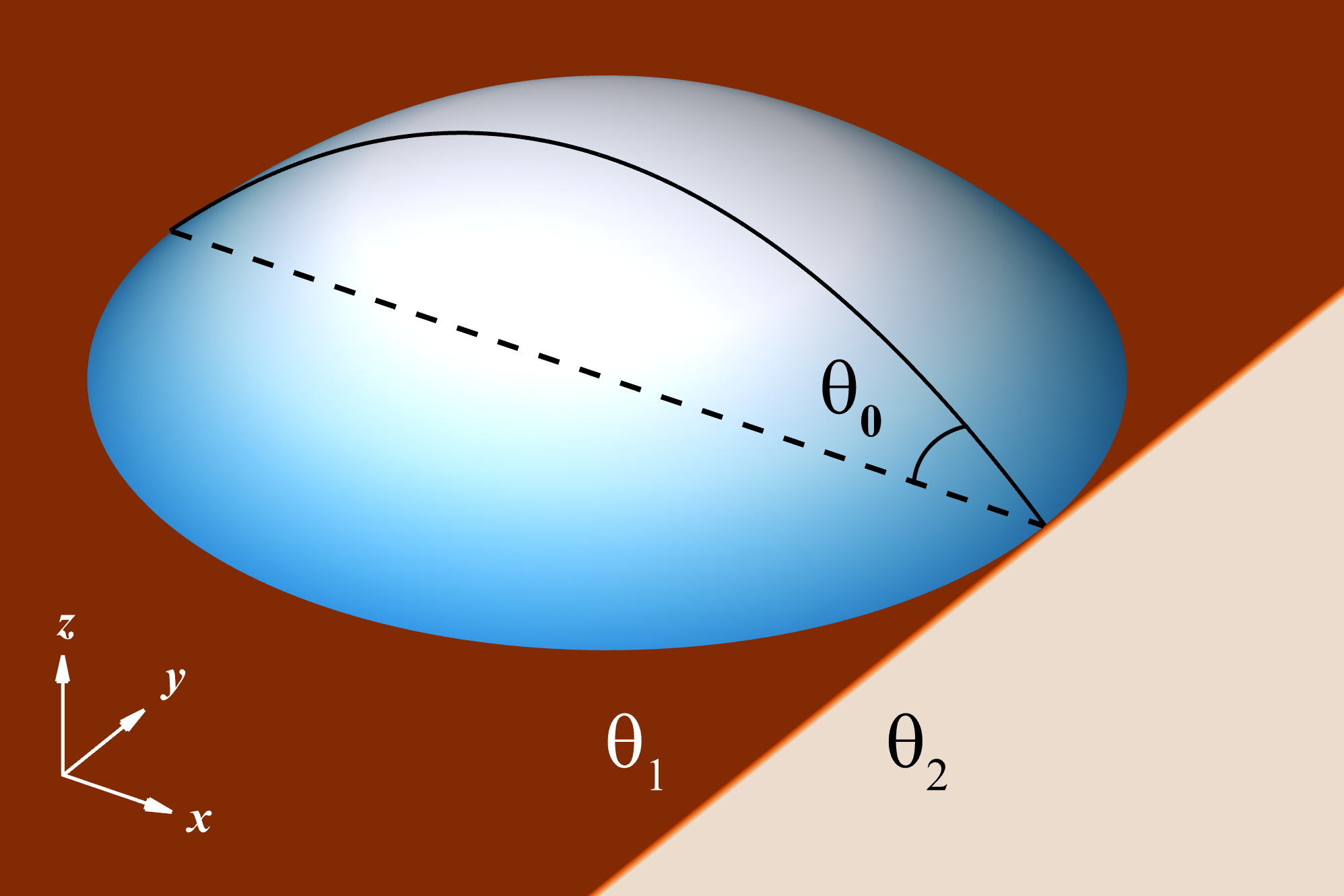}
    \caption{ Initial condition in the simulation of a 3D droplet driven by a chemical step. The chemical step is jointed by two homogeneous substrate. The left-right equilibrium contact angles $\theta_1$ and $\theta_2$ satisfy $\theta_1>\theta_2$.  The apparent contact angle of the droplet upon contact with the border is $\theta_0$.}
    \label{fig:sketch}
\end{figure}

Consider the dynamics of a droplet on a flat and horizontal substrate featuring a \textit{chemical step}, where the wettability distribution is prescribed by a step function of the microscopic contact angle. The wettability border is located at $x=0$, and the contact angle at $x<0$ and $x>0$ is denoted by $\theta_1$ and $\theta_2$, respectively. Without loss of generality, we assume $\theta_2<\theta_1$, i.e., the region of $x>0$ is more hydrophilic. Before contacting the wettability border, the droplet could be spreading. At the instant of contact, the apparent contact angle is $\theta_0$. We investigate the subsequent droplet motion towards the more hydrophilic region driven by wettability contrast. Both 2D and 3D situations are considered, with the  initial condition of the latter plotted in figure~\ref{fig:sketch}. The droplet is assumed to be of sufficiently small size and high viscosity such that the process is dominated by the balance between viscous and capillary forces, while inertial and gravitational effects can be neglected. Furthermore,  the contact angles $\theta_0$, $\theta_1$ and $\theta_2$ are assumed to be sufficiently small, enabling the employment of a lubrication equation of the droplet height $h(x, y, t)$. We adopt the Navier slip condition with a slip length $\lambda$ to alleviate the contact line singularity. 

The problem is nondimensionalized through the scalings
\begin{equation}
    (\bar x,\bar y)=\frac{1}{R}(x,y),\quad \bar h=\frac{h}{\theta_1 R},\quad \bar t=\frac{\gamma\theta_1^3}{3\mu R}t, \quad
    \bar\lambda=\frac{3\lambda}{\theta_1 R},  \label{nondim}
\end{equation}
where $R$ denotes the radius or half width of the wetted area of the droplet when it attains an equilibrium state on the less hydrophilic region. The nondimensional form of the governing equation and boundary conditions (BCs) with the bars omitted writes
\begin{equation}
    \partial_t h+\bnabla\bcdot\left[h^2(h+\lambda)\bnabla\nabla^2 h\right] =0,  \label{3DPDE}  
\end{equation}
\begin{equation}
    h=0,\quad \boldsymbol{n}\bcdot\bnabla h=-\Theta \quad \text{on}\quad  C.
\end{equation}
Here $\boldsymbol{n}$ is the outward unit normal and $C$ represents the domain boundary, which evolves with the contact line velocity
\begin{equation}
    \boldsymbol{U}|_{C}=\lambda h\boldsymbol{\nabla}\nabla^2h.
\end{equation}
The wettability distribution has the form
\begin{equation}
    \Theta=\begin{cases}
        1,&x<0,\\
        K,&x>0,
    \end{cases} \label{Theta}
\end{equation}
where $K={\theta_2}/{\theta_1}<1$ is the ratio of contact angles. At the wettability border $x=0$, the local contact angle may lie between $\theta_1$ and $\theta_2$ due to the contact line pinning, as will be discussed later. At the instant of contact with the border, the droplet macroscopically maintains a parabolic profile given by
\begin{equation}
    h=\left\{
    \begin{array}{ll}
      \frac{K_0}{2R_0}\left[R_0-(x+R_0)^2\right], & \mbox{for 2D}, \\[6pt]
      \frac{K_0}{2R_0}\left[R_0-(x+R_0)^2-y^2\right], & \mbox{for 3D},
  \end{array}\right.
  \label{IC}
\end{equation}
where $K_0=\theta_0/\theta_1$ is the rescaled apparent contact angle; $R_0$ is the instantaneous wetting radius or half width and can be obtained from volume constraint as $R_0=K_0^{1/2}$ and $K_0^{1/3}$ for 2D and 3D droplets, respectively. To guarantee a spreading droplet before contact with the border, we should have $K_0\ge1$; the situation of $K_0<1$ is beyond the scope of the present work since the corresponding droplet is expect to retract and does not interact with the border. We first investigate the case of $K_0=1$; the influence of different values of $K_0$ will be discussed in \S\,\ref{Sec:IC}, where it can be shown that only the early stage of evolution is affected. 

For millimetre-sized droplets of present interest, $\lambda\ll1$, thus numerically solving the full problem given by (\ref{3DPDE}) to (\ref{IC}) is challenging, owing to the presence of a numerical boundary layer near the contact line, which itself is moving. Moreover, the Navier slip condition does not fully resolve the contact line singularity, as the second and higher order derivatives of $h$ remain divergent at the contact line \citep{Buckingham2003}, which prevents direct numerical discretization of the original equation. In the present work, the full problem was solved numerically using a macroscopic algorithm recently developed by \cite{qin2024}. The core idea is to cut off the computational domain in the intermediate region near the contact line, where effective boundary conditions are derived from the local asymptotic solution of the film thickness. The cut-off reduces the computational cost, enabling simulations with a realistic slip length ($\lambda=3\times10^{-5}$ throughout the paper). This method requires a single-value dependence between the contact angle and the contact line velocity, which may be violated when contact line pinning occurs. To circumvent this constraint, the step function of the contact angle is smoothed as
\begin{equation}
    \Theta(x,y) = \frac{K-1}{2}\tanh\frac{x}{b}+\frac{K+1}{2},  \label{border}
\end{equation}
where $b\to0$ is the characteristic width of the smoothed wettability border, and it is fixed at $b=2.5\times10^{-3}$ throughout the paper. We have validated that halving the value of $b$ hardly affects the presented numerical results. 

\section{2D droplets} \label{Sec:2D}

\begin{figure}
    \centering
    \begin{minipage}{\textwidth}
        \centering
        \begin{overpic}[width=0.65\linewidth,trim=0cm 7cm 1.5cm 0.5cm,clip]{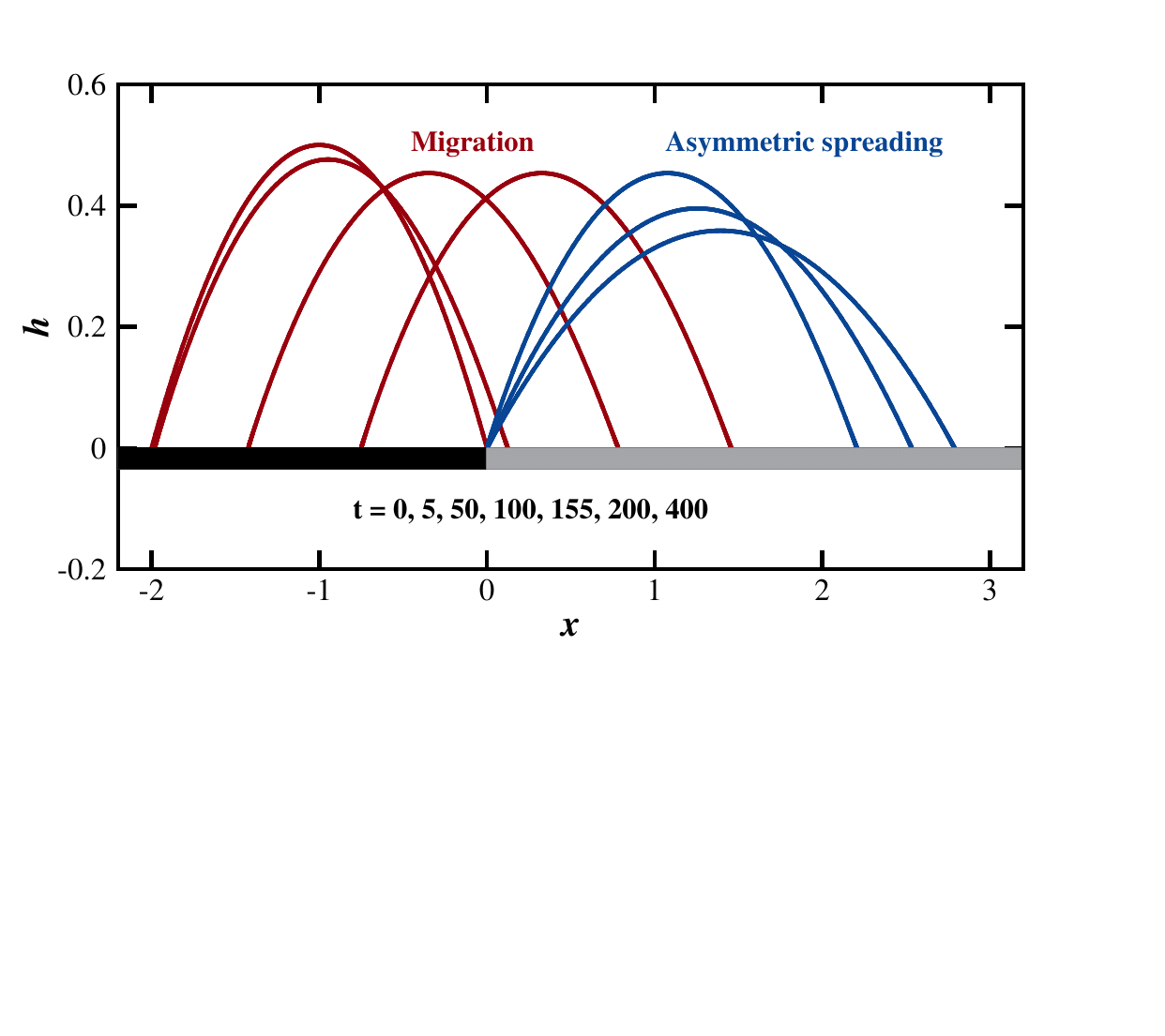}
            \put(-2,50){{\small (\textit{a})}}
        \end{overpic}
    \end{minipage}
    \\
    \begin{minipage}{0.48\linewidth}
        \centering
        \begin{overpic}[width=\linewidth,trim=0cm 0.8cm 1.5cm 3cm, clip]{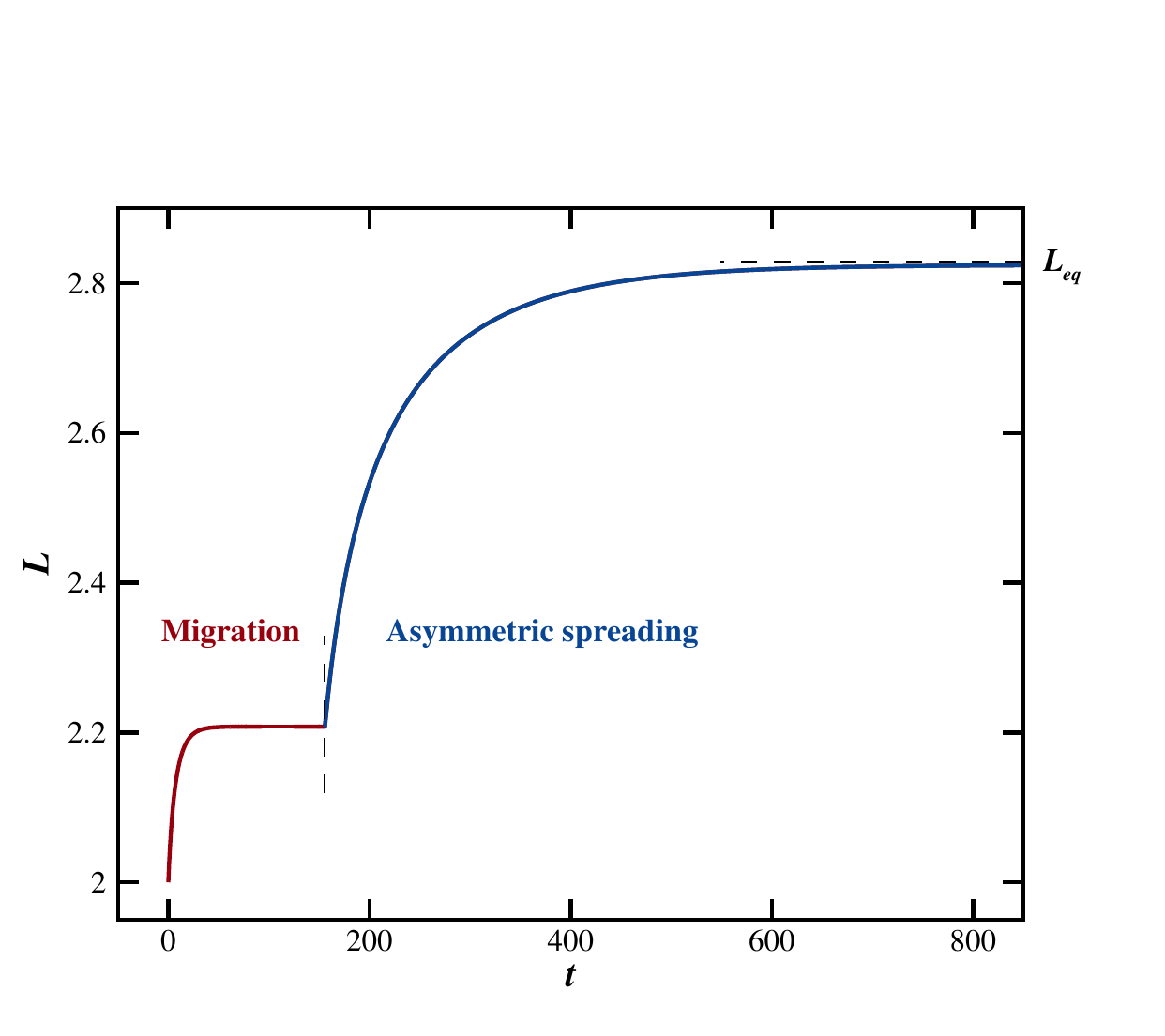}
            \put(-2,70){{\small (\textit{b})}}
        \end{overpic}
    \end{minipage}
    \hfill
    \begin{minipage}{0.48\linewidth}
        \centering
        \begin{overpic}[width=\linewidth,trim=0.1cm 0.8cm 1.5cm 3cm, clip]{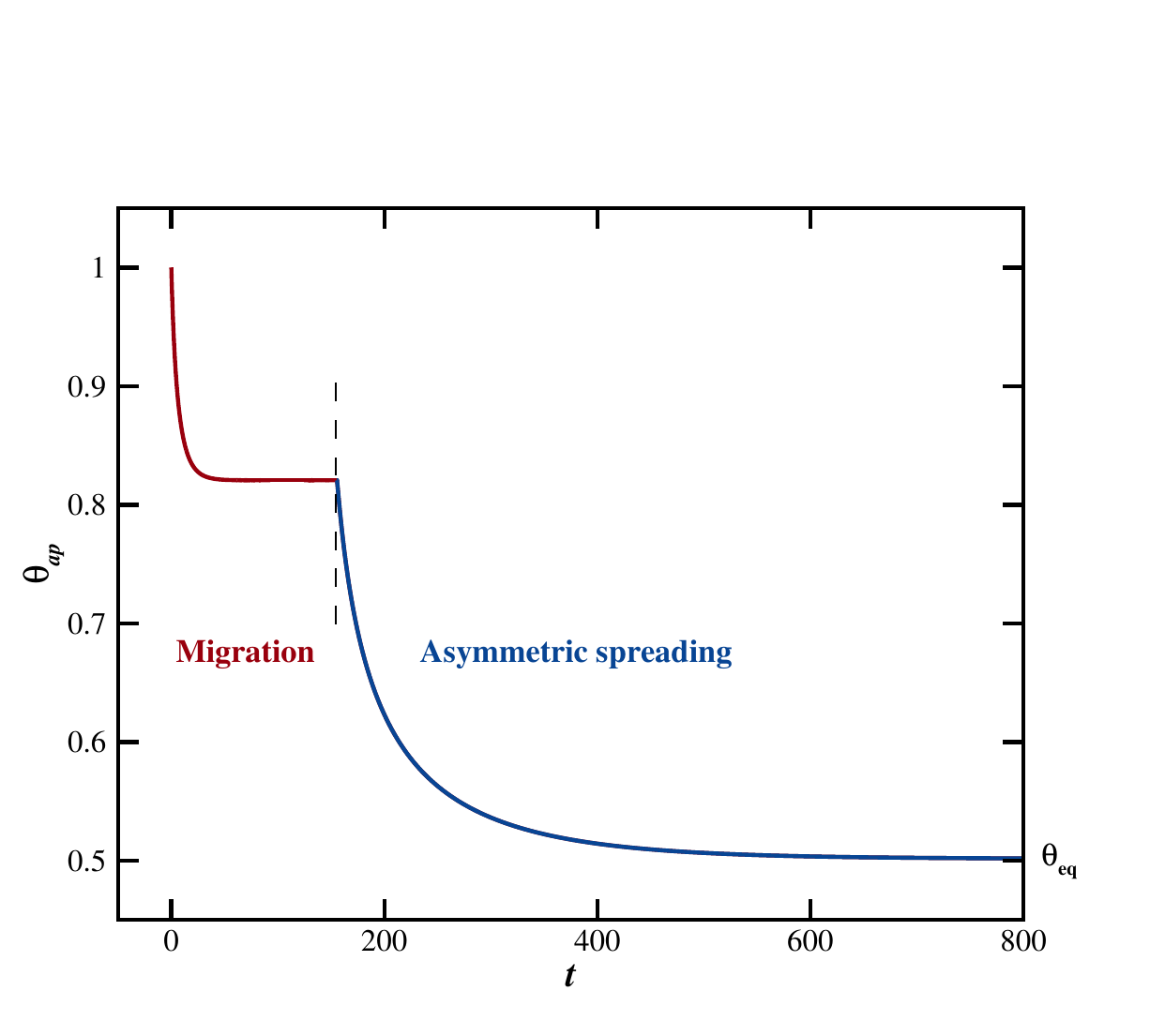}
            \put(-2,70){{\small (\textit{c})}}
        \end{overpic}
    \end{minipage}
    \caption{Evolution of a 2D droplet on a chemical step with $K=0.5$. (\textit{a}) Snapshots of droplet profiles. The evolution of (\textit{b}) droplet length $L$ and (\textit{c}) apparent contact angle $\theta_{ap}$ as functions of time.  At $t\approx 155$, marked by the vertical dashed line, the receding contact line reaches the wettability border at $x=0$, signifying the end of the migration stage and the onset of the asymmetric spreading stage. After  an early droplet elongation, $L$  and $\theta_{ap}$ remains constant during the migration stage. Then,  during the asymmetric spreading stage, $L$ and $\theta_{ap}$ approach the equilibrium value $L_{eq}=2/\sqrt{K}$ and $\theta_{eq}=K$ on the more hydrophilic region.}
    \label{fig:2D}
\end{figure}

We begin with an analysis of the 2D case, which is more tractable but still provides a qualitative representation of the 3D situation. The governing equations and BCs of a 2D droplet (or droplet stripe) on a chemical step reduce to
\begin{subequations}
    \begin{equation}
        \partial_th+\partial_x\left[h^2( h+\lambda  )\partial_x^3h\right] =0,\label{2Dpde}
    \end{equation}
    \begin{align}
        h=0,\quad \partial_xh=\Theta(x)\quad &\text{at}\quad x=x_1(t),\\
        h=0,\quad \partial_xh=-\Theta(x)\quad &\text{at}\quad x=x_2(t),
    \end{align}
    \begin{equation}
        \dot x_1=\lambda h\partial_x^3h|_{x_1},\quad \dot x_2=\lambda h\partial_x^3h|_{x_2},
    \end{equation}
\end{subequations}
where $x_1$ and $x_2$ denote the position of the front and rear contact line, respectively.

It can be expected that the droplet moves from the initial profile given by \eqref{IC} to an equilibrium state on the more hydrophilic region. The evolution of the droplet profile is illustrated in figure~\ref{fig:2D}(\textit{a}) for $K=0.5$. After an initial transient period of stretching, the droplet migrates towards the more hydrophilic side. At $t\approx155$, the receding contact line reaches the wettability border and remains pinned thereafter, while the other contact line continues to advance and thereby elongating the droplet. As depicted in figure~\ref{fig:2D}(\textit{b}), the droplet length, defined as $L\equiv x_2-x_1$, after a brief rise, remains constant for a period, corresponding to a steady displacement of the droplet. Then $L$ increases again until it eventually approaches the equilibrium length at the more hydrophilic substrate, $L_{eq}=2/\sqrt{K}$.  Correspondingly, the evolution of the droplet apparent contact angle $\theta=4/L^2$, defined by fitting the droplet profile as a parabola, is shown in figure~\ref{fig:2D}(\textit{c}). We can thus identify two distinct stages: 1) the {\it migration} stage, when the droplet traverses across the wettability border from the less to the more hydrophilic side; 2) the {\it asymmetric spreading} stage, when the droplet spreads on the more hydrophilic side with one contact line pinned at the wettability border.  Because of the smoothing \eqref{border}, the pinned contact line is modelled by a moving one with sufficiently low speed, which is apparently pinned on the macroscopic scale. As introduced in the following sections, it turns out that analytical solutions can be obtained by considering these two stages separately, and the smoothing can be discarded with additional treatment to investigate a strictly pinned contact line.

\subsection{Migration}\label{Subsec:Mig}
 The migration of the droplet is driven by the wettability contrast and can be described by the asymptotic theory of \citet{vellingiri2011}, who considered a continuous wettability distribution. We focus here on the steady motion of the droplet during the migration stage, in which both the width $L$ and the droplet velocity $\delta$ are constant, allowing for more explicit solutions. We also consider the special case of $K=0$, which has not been investigated in \citet{vellingiri2011}. It is convenient to introduce a reference frame attached to the droplet. Thus the governing equation (\ref{2Dpde}), after one integral, yields
\begin{equation}
    (h^2+\lambda h)h'''=\delta, \label{migODE}
\end{equation}
where the primes denote derivatives with respect to $x$. The BCs are
\begin{subequations}
    \begin{equation}
        h=0,\quad h'=1 \quad \text{at}\quad  x=0,
    \end{equation}
    \begin{equation}
        h=0,\quad h'=-K \quad \text{at}\quad  x=L.
    \end{equation}    \label{migBC}
\end{subequations}
The third-order boundary value problem with two undetermined parameter $L$ and $\delta$ can be closed with the volume conservation condition
\begin{equation}
    \int_{0}^{L}h\,\mathrm{d}x=\frac{2}{3}, \label{volCons}
\end{equation}
where the value is specified by the initial profile \eqref{IC}.

It is well-established that such a system, involving moving contact lines, is subject to multi-scale solutions with boundary layers near the contact line. When $\lambda\ll1$, the droplet has an outer region with negligible slip effect and two inner regions of thickness $\mathcal{O}(\lambda)$ near the contact lines, where slip effect is significant. Following \cite{hocking1983}, we present asymptotic solutions of these distinct regions in the limit of slow migration ($\delta\to0$). The matching of these solutions determines the droplet profile and velocity.

Omitting the slip term $\lambda h$ in (\ref{migODE}), the outer solution describing the macroscopic droplet profile satisfies
\begin{subequations}
    \begin{gather}
        h^2h'''=\delta, \\
        h=0 \quad \text{at}\quad x=0,\ L,    \\
        \int_{0}^{L}h\, \mathrm{d}x=\frac{2}{3},
    \end{gather}
    \label{outEq}
\end{subequations}
where the contact angle conditions are unnecessary. The outer equation (\ref{outEq}\textit{a}) admits an analytical solution expressed by Airy functions \citep{Duffy1997}, which was adopted by \cite{eggers2004,eggers2005} and \cite{Snoeijer2010} in studying forced dewetting and, more relevantly to the present problem, by \cite{pismen2006} for a moving droplet driven by a wettability gradient. In the present work, we alternatively employ the classical asymptotic procedure developed by \citet{hocking1983} for advancing contact lines. In this procedure, the outer and inner solutions are matched through a Voinov region where the cube of the film slope varies logarithmically \citep{voinov1976}. For receding contact lines, as demonstrated by \cite{Duffy1997}, the Voinov region still exists and was used by \cite{eggers2005} to construct an inner solution that can be directly matched to the outer solution. For the present problem, we intended to match the inner and outer solutions in the Voinov region for both advancing and receding contact lines, which is feasible for slow migration; the corresponding matching results agree perfectly with the numerical results, as shown later. This method relies on an expansion of the film thickness in terms of small contact line speed, i.e., $h=h_0+\delta h_1+\mathcal{O}(\delta^2)$. Substituting into \eqref{outEq} and solving sequentially for the leading-order solution $h_0$ and the first-order solution $h_1$ provides the perturbation solution for the outer region. 

The leading-order approximation is simply a hydrostatic profile that writes
\begin{equation}
    h_0=\frac{4}{L^3}x(L-x),    \label{h0}
\end{equation}
corresponding to an apparent contact angle $\theta_{ap}=4/L^2$. The first-order equation and supplementary conditions are
\begin{subequations}
    \begin{gather}
        h_0^2h_1'''=1,   \\
        h_1=0 \quad \text{at}\quad x=0,\ L,  \\
        \int_{0}^{L}h_1\, \mathrm{d}x=0,
    \end{gather}
\end{subequations}
and the solution can be obtained as
\begin{equation}
    h_1=\frac{L^3}{16}\left[x(L-x)\ln\frac{L-x}{x}\right].
\end{equation}
The matching with the inner region can be directly accomplished by the cube of the slope without invoking an intermediate region \citep{eggers2005, sibley2015}. Thus the outer matching conditions are
\begin{align}
    \left(h_{out}'\right)^3 &\sim \left(\frac{4}{L^2}\right)^3-3\delta\ln\frac{\mathrm{e}x}{L} \quad \text{as}\quad x\to0,  \label{migOutRe}    \\
    \left(h_{out}'\right)^3 &\sim -\left(\frac{4}{L^2}\right)^3-3\delta\ln\frac{\mathrm{e}(L-x)}{L} \quad \text{as}\quad x\to L.    \label{migOutAd}
\end{align}

In the inner region, the slip is significant and the full equation near the two contact lines should be solved separately. Near the advancing contact line, introduce the inner variables $H(\xi)=h/\lambda$ and $\xi=(L-x)/\lambda$ and the problem is transformed into
\begin{subequations}
    \begin{gather}
        (H^2+H)H'''=-\delta, \\
        H(0)=0,\quad  H'(0)=K, \quad H''(\infty)=0.
    \end{gather}
    \label{eq:inner}
\end{subequations}
The asymptotic solution of \eqref{eq:inner} for small $\delta$, or more strictly speaking $\delta/K^3\ll1$, is well documented in the literature \citep{hocking1983, eggers2005,sibley2015}. For the present purposes, the far-field asymptotics of the inner solution expressed in the outer variables writes
\begin{equation}
    \left(h_{in}'\right)^3 \sim -K^3-3\delta\ln{\frac{\mathrm{e}K(L-x)}{\lambda}} \quad \text{as}\quad L-x\gg\lambda. \label{migInAd}
\end{equation}
Similarly, the inner condition at the receding contact line is
\begin{equation}
    \left(h_{in}'\right)^3 \sim 1-3\delta\ln{\frac{\mathrm{e}x}{\lambda}} \quad \text{as}\quad  x\gg\lambda. \label{migInRe}
\end{equation}

Matching (\ref{migOutRe}) with (\ref{migInRe}), and (\ref{migOutAd}) with (\ref{migInAd}), yields
\begin{align}
    \left(\frac{L}{2}\right)^6&=\frac{\ln K+2\ln(L/\lambda)}{\ln K+(K^3+1)\ln(L/\lambda)},  \label{finKL}\\
    \delta &= \frac{1-(2/L)^6}{3\ln(L/\lambda)}.  \label{finKDel}
\end{align}
The velocity only depends explicitly on the droplet length. One can substitute \eqref{finKL} into \eqref{finKDel} to obtain
\begin{equation}
    \delta = \frac{1-K^3}{3[\ln K+2\ln(L/\lambda)]},
\end{equation}
which indicates that $\delta$ is proportional to $1-K^3$ up to a logarithmic correction. Moreover, in the limit of $K\to1$, we have $L-2=(1-K)/2+\mathcal{O}\left[(1-K)^2\right]$ according to \eqref{finKL}. 

It is worth noting that (\ref{migInAd}) conforms with the expansion only when $K^3\gg\delta$. For conditions that $K$ is too small, the theory no longer holds. For instance, the expression (\ref{finKL}) predicts an infinite droplet length as $K\to0$, which is obviously unphysical. For small values of $K$, the theory can be modified by replacing the inner condition at the advancing contact line (\ref{migInAd}) with
\begin{equation}
    \left(h'_{in}\right)^ 3\sim -q\delta-3\delta\ln{\frac{\mathrm{e}\delta^{1/3}(L-x)}{\lambda}} \quad \text{as}\quad L-x\gg\lambda,
    \label{eq:insk}
\end{equation}
where $q=0.74+K^2\delta^{-2/3}$ is an approximate relation proposed by \cite{hocking1992}. Matching at the advancing contact line gives
\begin{equation}
    \left(\frac{2}{L}\right)^6-\delta\left(q+3\ln\frac{L\delta^{1/3}}{\lambda}\right)=0, \label{smaKmatch}
\end{equation}
which, combined with the other condition \eqref{finKDel}, determines $L$ and $\delta$ for given values of $K$. Note that $K$ exists explicitly in the expression of $q$ only.

\begin{figure}
    \centering
    \begin{minipage}{.45\textwidth}
        \begin{overpic}[width=\linewidth,trim=1cm 0.2cm 2cm 1cm,clip]{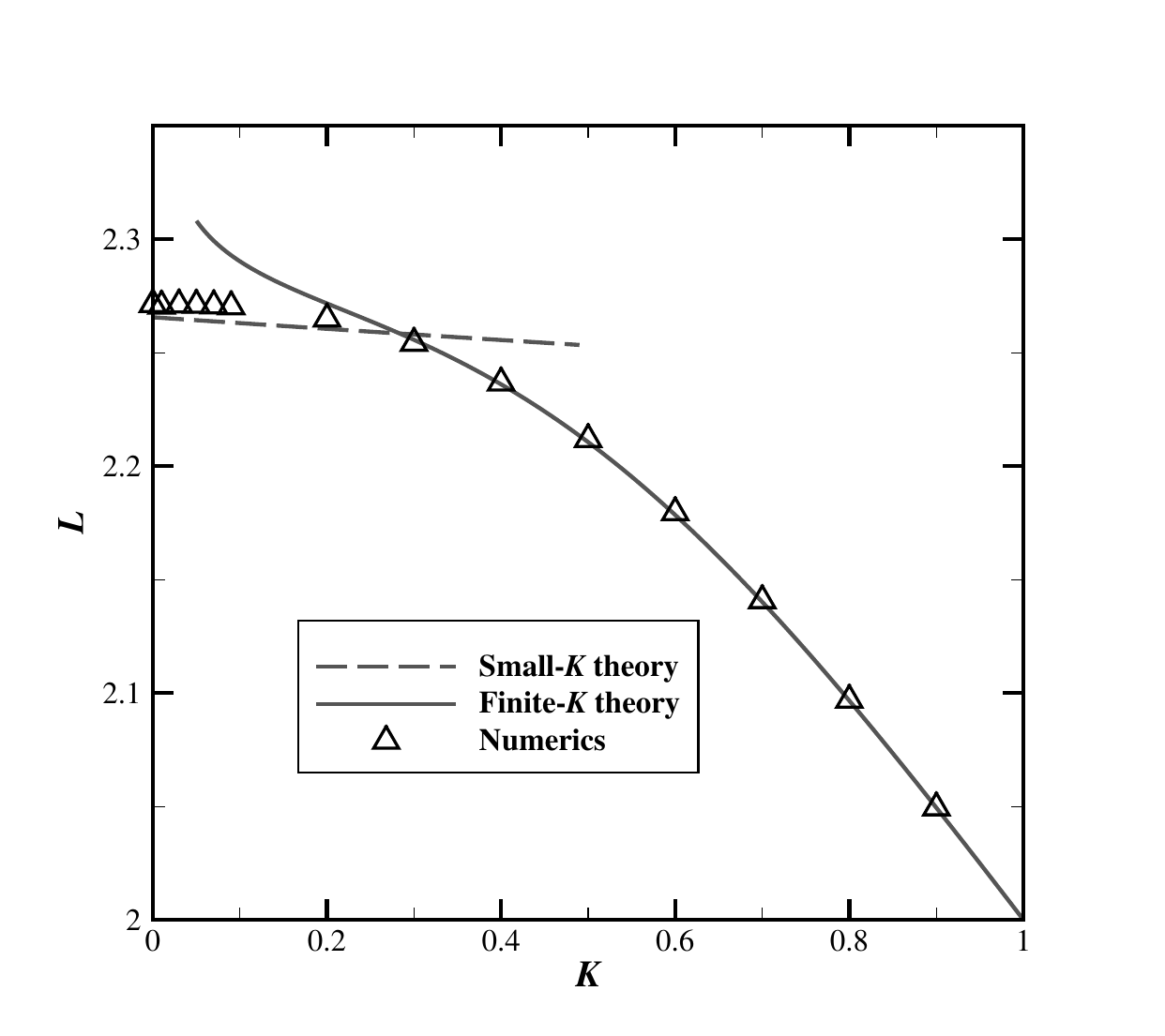}
            \put(-5,87){{\small (\textit{a})}}
        \end{overpic}
    \end{minipage}
    \hfill
    \begin{minipage}{0.45\linewidth}
        \begin{overpic}[width=\linewidth,trim=0.5cm 0.2cm 2cm 1cm,clip]{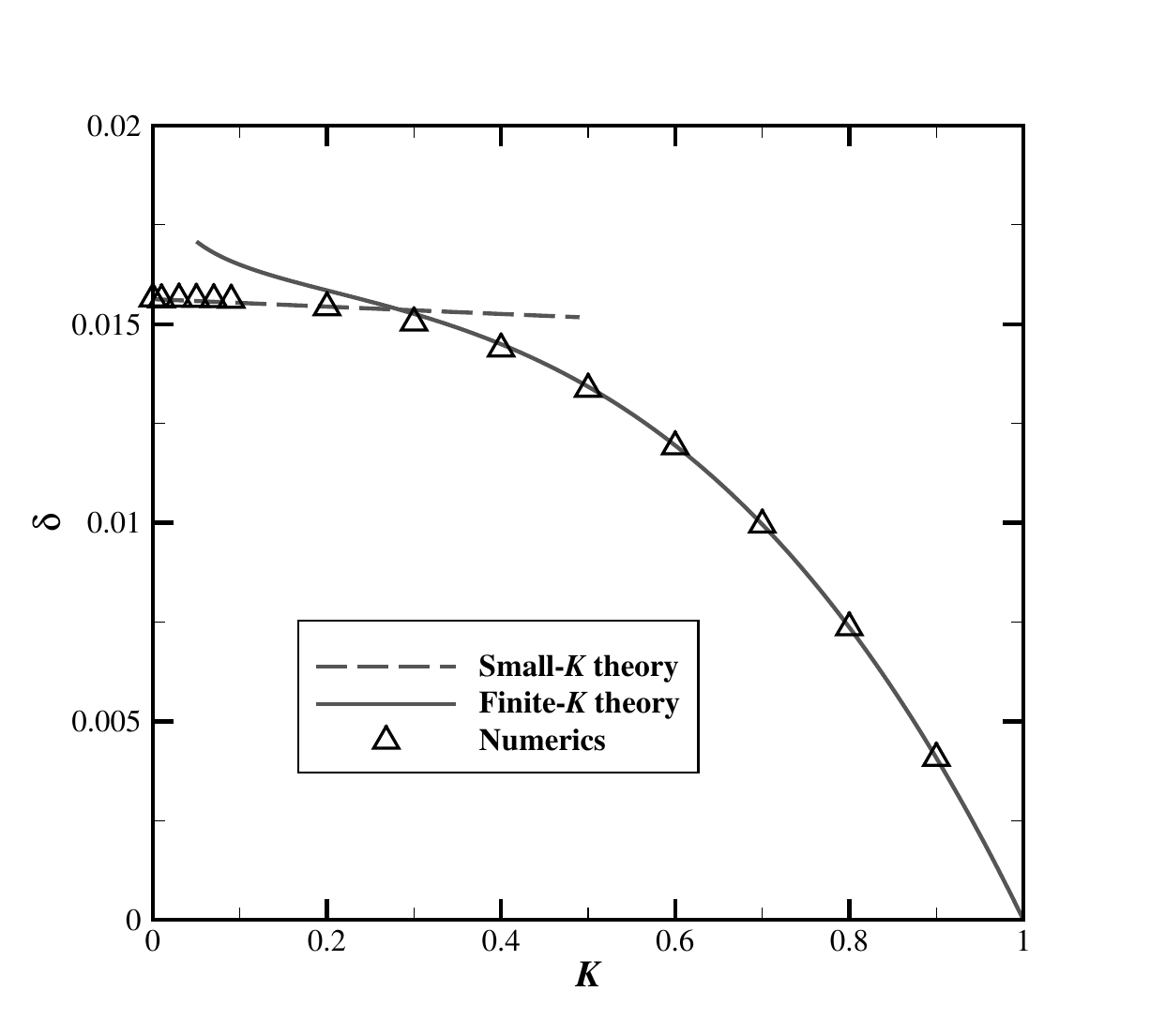}
            \put(-5,87){{\small (\textit{b})}}
        \end{overpic}
    \end{minipage}
    \\
    \begin{minipage}{0.45\linewidth}
        \begin{overpic}[width=\linewidth,trim=1cm 0.2cm 2cm 1cm,clip]{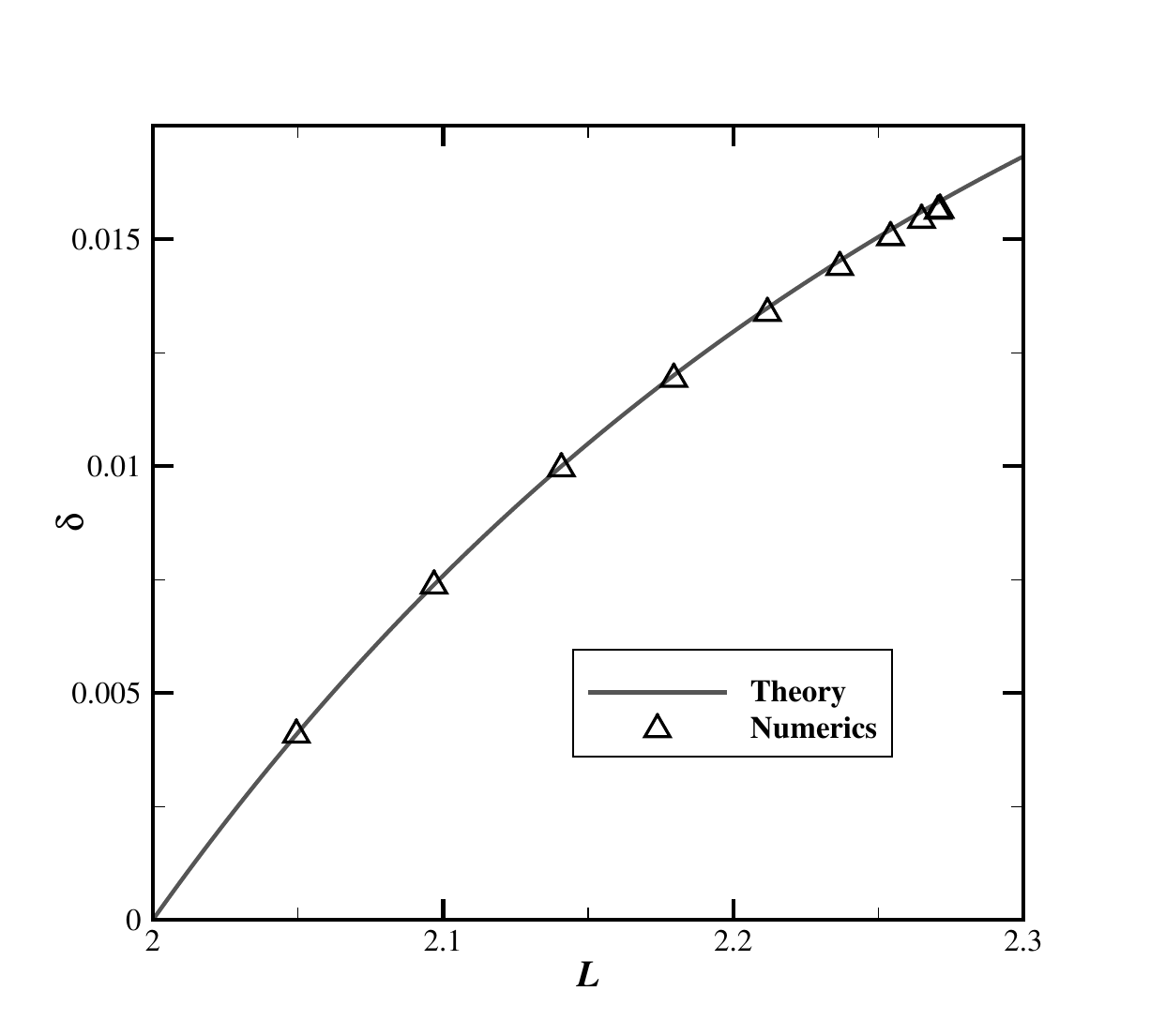}
            \put(-5,87){{\small (\textit{c})}}
        \end{overpic}
    \end{minipage}
    \hfill
    \caption{Comparison of the theoretical and numerical results during the steady migration stage. Variation of (\textit{a}) the droplet length $L$ and (\textit{b}) the migration velocity $\delta$ as a function of $K$. The finite-$K$ theory corresponds to \eqref{finKL} and \eqref{finKDel}; the small-$K$ theory corresponds to \eqref{finKDel} and \eqref{smaKmatch}.  (\textit{c}) Migration velocity $\delta$ as a function to $L$ with the theoretical curve given by \eqref{finKDel}. 
    }
    \label{fig:mig}
\end{figure}

Figure~\ref{fig:mig} compares the finite-$K$ and the small-$K$ theories with numerical results of (\ref{migODE}) and (\ref{migBC}), which is obtained using the numerical approach described in Appendix~\ref{appC}. As expected, the droplet becomes longer and moves faster at smaller $K$ since the driving force becomes stronger. The finite-$K$ theory shows excellent agreement with the numerical results for $K\gtrsim0.2$. For $K<0.2$, both $K$ and $\delta$ remain finite and are reasonably predicted by the small-$K$ theory, while the finite-$K$ theory fails. The dependence between $\delta$ and $L$ is shown in figure~\ref{fig:mig}(\textit{c}), in which the solid curve represents \eqref{finKDel}, valid for both finite and small values of $K$ as confirmed by the numerical solutions.

\subsection{Asymmetric spreading}
The migration stage ends when the receding contact line reaches the wettability border and hence all the liquid is accumulated on the more hydrophilic region of the substrate. At this instant, the apparent contact angle $\theta_0=4/L^2$ lies within the interval $(K,1)$. Therefore, the droplet has yet to reach its equilibrium state on the more hydrophilic region. A further spreading process ensues until the apparent contact angle decreases to $K$. During this stage, the advancing contact line keeps moving and the other contact line is pinned at the border, leading to an asymmetric spreading.

The dynamics of the asymmetric spreading stage is governed by (\ref{2Dpde}), while the boundary conditions requires appropriate treatment due to the presence of the pinned contact line. Different from the moving contact line, the contact angle is not known \textit{a priori} at the pinned contact line and hence cannot be implemented as a boundary condition. Instead, a zero-flux condition $h^2\partial_x^3h=0$ is adopted, and the BCs write
\refstepcounter{equation}\label{sprBC}
$$
  h=0,\quad h^2\partial_x^3h=0 \quad \text{at}\quad  x=0,
  \eqno{(\theequation{\mathit{a},\mathit{b}})}
$$
$$
  h=0,\quad \partial_xh=-K \quad \text{at}\quad  x=L.
  \eqno{(\theequation{\mathit{c},\mathit{d}})}
$$
The contact line is therefore strictly pinned at $x=0$ and the smoothed border \eqref{border} can be discarded. The volume conservation condition \eqref{volCons} can be used to fully determine the solution.

An asymptotic analysis can be performed as well when $\lambda\ll1$. It turns out that a boundary layer still occurs at the pinned contact line. The whole drop can thus be divided into an outer region and two inner regions near the contact lines. In the outer region, the slip term and the contact angle condition are neglected. For small spreading rate $\dot{L}$, we seek the outer solution of the form
\begin{equation}
    h(x,t)=h_0(x,L)+\dot Lh_1(x,L)+\mathcal{O}(\dot L^2).
    \label{eq:hpin}
\end{equation}
Accordingly, we have $\partial_t h=\dot L\partial_L h_0+\mathcal{O}(\dot{L}^2)$. The leading-order governing equation and BCs are obtained as
\begin{subequations}
    \begin{gather}
        \partial_x\left(h_0^3\partial_{x}^{3}h_0\right)=0, \label{sprh0eq}   \\
        h_0=0,\quad h_0^2\partial_x^3h_0=0 \quad \text{at}\quad  x=0, \label{sprh0BC}  \\
        h_0=0 \quad \text{at}\quad  x=L, \\
        \int_{0}^{L}h_0\, \mathrm{d}x=\frac{2}{3}.
    \end{gather}
\end{subequations}
Integrating (\ref{sprh0eq}) once and using (\ref{sprh0BC}) yield $\partial_{x}^{3}h_0=0$. Thus the leading-order solution represents again a hydrostatic profile:
\begin{equation}
    h_0=\frac{4}{L^3}x(L-x).    \label{sprh0}
\end{equation}
The first-order solution $h_1$ satisfies
\begin{subequations}
    \begin{gather}
        \partial_L h_0+\partial_x\left(h_0^3\partial_x^3 h_1\right)=0.\\
        h_1=0,\quad 2h_0h_1\partial_x^3h_0+h_0^2\partial_x^3h_1=0 \quad \text{at}\quad  x=0, \\
        h_1=0 \quad \text{at}\quad  x=L, \\
        \int_{0}^{L}h_1\, \mathrm{d}x=0.
    \end{gather}
\end{subequations}
The solution can be found as
\begin{equation}
    h_1=\frac{L^3}{32}\left[3x(L-x)+x^2\ln\frac{x}{L}+(L^2-x^2)\ln\left(\frac{L-x}{L}\right)\right].    \label{sprh1}
\end{equation}
The first-order derivative of $h_1$ diverges at $x=L$, indicating the well-known singularity of moving contact lines. At the pinned contact line, $h_1$ has a finite slope, which can serve as a correction to the definition of the apparent contact angle,
\begin{equation}
    \theta=\partial_xh_0(0)+\dot{L}\partial_xh_1(0)=\frac{4}{L^2}+\frac{L^4\dot L}{16}.
    \label{eq:AppAnglePin}
\end{equation}
Interestingly, the second-order derivative still exhibits a logarithmic singularity, indicating the presence of a boundary layer at the pinned contact line.

The inner solution at the advancing contact line is given by \eqref{migInAd} for finite $K$ and by \eqref{eq:insk} for small $K$, with contact-line speed $\delta$ replaced by $\dot{L}$. Matching these solutions with the cube of the derivative of the outer solution \eqref{eq:hpin}, we arrive at a reduced equation of the droplet width $L$ which, for finite $K$, reads
\begin{equation}
    \dot L=\frac{(2/L)^6-K^3}{3\ln(KL/\mathrm{e}\lambda)}. \label{dotL}
\end{equation}
As the equilibrium state is approached, it has $\dot{L}\to0$ and $L\to L_{eq}=2/\sqrt{K}$. (\ref{dotL}) can be easily solved and validated against the numerical solution (which is introduced in Appendix~\ref{appD}), as shown in figure~\ref{fig:sprL_t}, where the initial value of $L$ is given by the steady migration stage. The results show that a smaller value of $K$ leads to a faster contact line velocity, but prolongs the time to reach equilibrium. For small $K\to0$, the evolution of the droplet length is given by
\begin{equation}
    \dot L\left(q+3\ln\frac{L\dot L^{1/3}}{\mathrm{e}\lambda}\right)-\left(\frac{2}{L}\right)^6=0,
    \label{eq:smallKpin}
\end{equation}
where $q=0.74+K^2\dot L^{-2/3}$. This relation is used to produce the theoretical curve of $K=0$ in figure~\ref{fig:sprL_t}. In this special case of complete wetting, the droplet spreads infinitely as no equilibrium state is expected. For small yet finite values of $K$, we note that \eqref{eq:smallKpin} does not hold when approaching the equilibrium state, because the contact line speed $\dot{L}\to0$, thereby violating the assumption $\dot{L}/K^3\gg1$ required by \eqref{eq:insk}. In this situation, one can adopt \eqref{dotL} to describe the very late stage of spreading, which, however, is out of the scope of the present work.

\begin{figure}
    \centering
    \includegraphics[width=.6\linewidth,trim=1cm 0.5cm 1.5cm 1cm,clip]{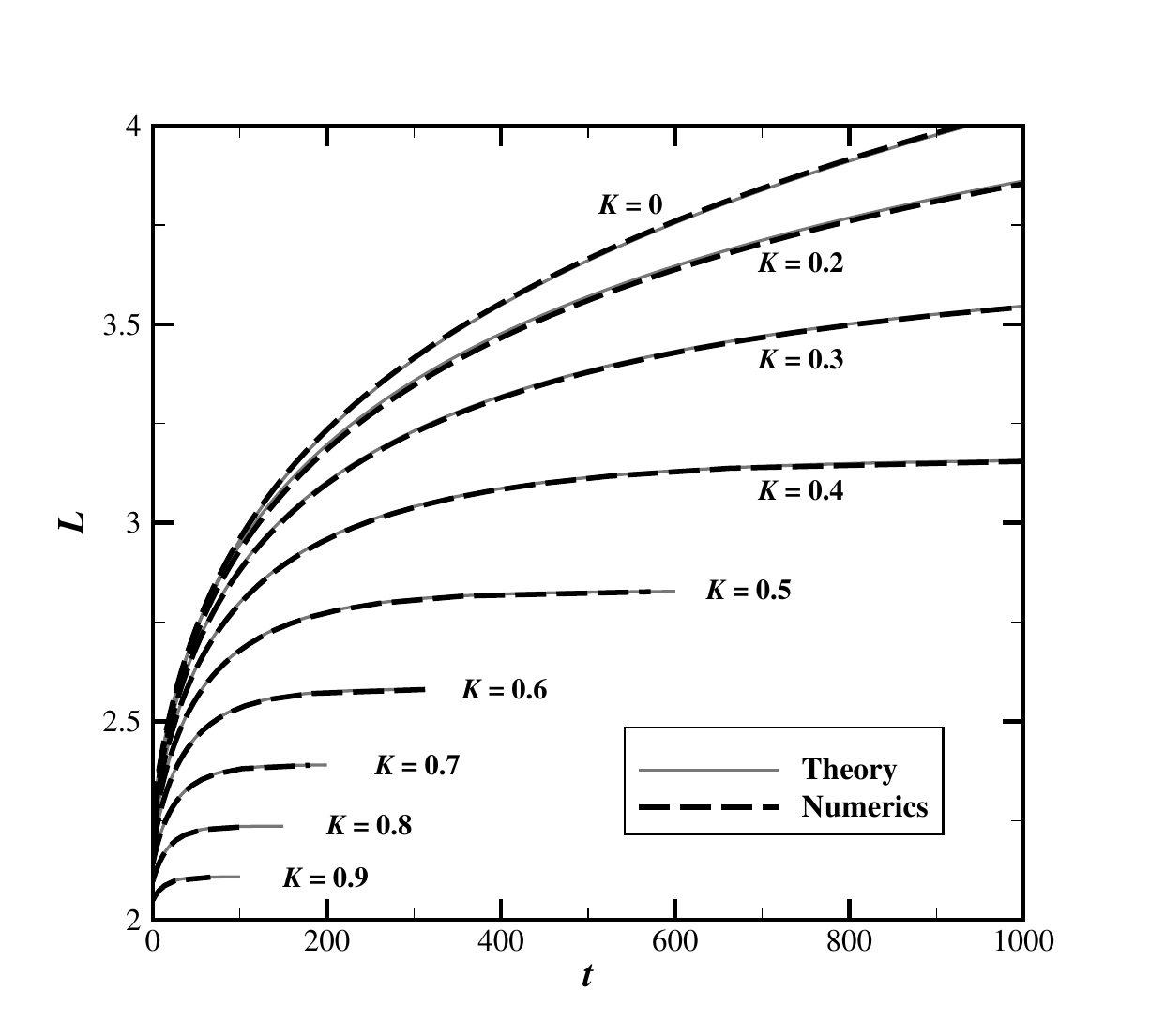}
    \caption{Temporal evolution of the droplet length $L$ during the asymmetric spreading stage.}
    \label{fig:sprL_t}
\end{figure}

Besides the conditions of vanishing film thickness and flux, the derivation of the droplet evolution does not require any knowledge of the microscopic behaviour near the pinned contact line, implying that the local behaviour near the pinned contact line is determined by the macroscopic dynamics. Accordingly, the outer solution supplies a far-field condition for the inner region. We present an asymptotic analysis to elucidate the structure of the boundary layer near the pinned contact line, and demonstrate that the curvature singularity can be regularized by introducing the Navier slip. To this end, we introduce the inner variables
\begin{equation}
    \xi=\frac{x}{\lambda},\quad H=\frac{h}{\lambda}.
\end{equation}
The governing equation for the inner region becomes
\begin{equation}
    \lambda\partial_t H+\partial_\xi \left[H^2(H+1)\partial_\xi^3H\right]=0,
    \label{eq:AsyInner}
\end{equation}
with BCs
\begin{equation}
    H=0,\quad H^2\partial_\xi^3H=0,\quad \text{at}\quad \xi=0.
\end{equation}
As $\xi\to\infty$, we anticpate that the inner and outer solutions can be  matched. Expressing \eqref{eq:hpin} together with \eqref{sprh0} and \eqref{sprh1} in terms of the inner variables and expanding it for large $\xi$, we obtain 
\begin{equation}
    H\sim\frac{4\xi}{L^2}+\frac{L^4\xi}{16}\dot L-\frac{4\xi^2}{L^3}\lambda+\frac{L^3\xi^2}{32}\epsilon+\frac{L^3\xi^2}{64}\left(2\ln\frac{\xi}{L}-7\right)\dot L\lambda+\cdots,\quad \text{as}\quad \xi\to\infty, \label{Hinfty}
\end{equation}
where $\epsilon\equiv\dot L\lambda\ln\lambda$. This asymptotic relation serves as the far-field condition for the inner region.

Inspired by the form of \eqref{Hinfty}, we consider a more general situation in which $\dot L$, $\lambda$ and $\epsilon$ are independent and vanishingly small, and seek an inner expansion in the form
\begin{equation}
    H(\xi,t)=H_0(\xi,L)+\dot LH_1(\xi,L)+\lambda H_2(\xi,L)+\epsilon H_3(\xi,L)+\dot L\lambda H_4(\xi,L)+\cdots.
\end{equation}
Then the first term in \eqref{eq:AsyInner} becomes
\begin{equation}
    \lambda\partial_t H=\dot L\lambda \partial_L H_0+\cdots,
\end{equation}
which allows the time derivative to be dropped up to order $\mathcal{O}(\dot L\lambda)$. It is straightforward to successively demonstrate that 
\begin{equation}
     \partial_\xi\left[H_0^2(H_0+1)\partial_\xi^3H_n\right]=0,
\end{equation}
for $n\in\{0,1,2,3\}$, and
\begin{equation}
    \partial_LH_0+\partial_\xi\left[H_0^2(H_0+1)\partial_\xi^3H_4\right]=0.
\end{equation}
The BCs are obtained as
\begin{equation}
     H_n=0,\quad H_0^2\partial_\xi^3H_n=0,\quad \text{at}\quad \xi=0,   \label{Hn0}
\end{equation}
for $n\in\{0,1,2,3,4\}$, and
\begin{equation}
     H_0\sim \frac{4}{L^2}\xi, \quad  H_1\sim \frac{L^4}{16}\xi,\quad H_2\sim-\frac{4}{L^3}\xi^2 ,\quad H_3\sim\frac{L^3}{32}\xi^2 ,\quad H_4\sim\frac{L^3\xi^2}{64}\left(2\ln\frac{\xi}{L}-7\right),
\end{equation}
as $\xi\to\infty$. The corresponding solutions can be found as
\begin{gather}
    \begin{split}
    H_0=\frac{4}{L^2}\xi, \quad H_1=\frac{L^4}{16}\xi,\quad 
    H_2=-\frac{4}{L^3}\xi^2+A\xi,\quad 
    H_3=\frac{L^3}{32}\xi^2+B\xi,\\
    H_4=\frac{L^3}{512}\left[(L^2+4\xi)^2\ln(L^2+4\xi)-2L^4\ln{L}\right]-\frac{2\ln(4L)+7}{64}L^3\xi^2+C\xi.
    \end{split}
    \label{asymSprH}
\end{gather}
The coefficients $A$, $B$ and $C$ remain to be determined; however, the corresponding terms only introduce an $\mathcal{O}(\lambda)$ correction of the film thickness and slope, and do not contribute to the curvature. Accordingly, the surface slope in terms of outer variables, after assigning $\epsilon=\dot L\lambda\ln\lambda$, is given by
\begin{equation}
  \partial_x h=\frac{4}{L^2}+\frac{L^4\dot L}{16}+\mathcal{O}(\lambda),  \label{thetaTheo}
\end{equation}
which is consistent with \eqref{eq:AppAnglePin} and hardly varies across the inner region near the pinned contact line. The second-order derivative has the form
\begin{equation}
    \partial^2_x h=\kappa_{ap}+\frac{L^3\dot L}{16} \ln\frac{\lambda L^2+4x}{4\mathrm{e}^2L},
    \label{eq:curvature}
\end{equation}
where $\kappa_{ap}=-{8}/{L^3}$ is the curvature of the leading-order approximation of the outer solution \eqref{sprh0}, namely the apparent curvature. In particular, the curvature has been regularized in the presence of slip, leading to a finite microscopic curvature at the contact line
\begin{equation}
    \kappa_m=\kappa_{ap}+\frac{L^3\dot L}{16} \ln\frac{\lambda L}{4\mathrm{e}^2},
\end{equation}
which is a counterpart of the well-known Cox--Voinov law for moving contact lines \citep{voinov1976,cox1986}.

A comparison between the asymptotic theory and numerical results is presented in figure~\ref{fig:pinInner}; the latter are obtained using the numerical method described in Appendix~\ref{appD}. As shown in figure~\ref{fig:pinInner}(\textit{a}), the temporal decrease of the contact angle during spreading is accurately predicted by (\ref{thetaTheo}). The instantaneous profile of the film curvature is plotted in figure~\ref{fig:pinInner}(\textit{b}). It is shown that the outer solution well predicts the macroscopic variation of the curvature but fails to capture the finite curvature upon approaching the pinned contact line within a distance of $\mathcal{O}(10\lambda)$, where the inner solution is necessary to describe the microscopic curvature. The outer and inner solutions overlap in a region characterized by a logarithmic variation of the curvature.

\begin{figure}
    \centering
    \begin{minipage}{.45\textwidth}
    \begin{overpic}[width=\linewidth,trim=0.5cm 0.2cm 2cm 1cm,clip]{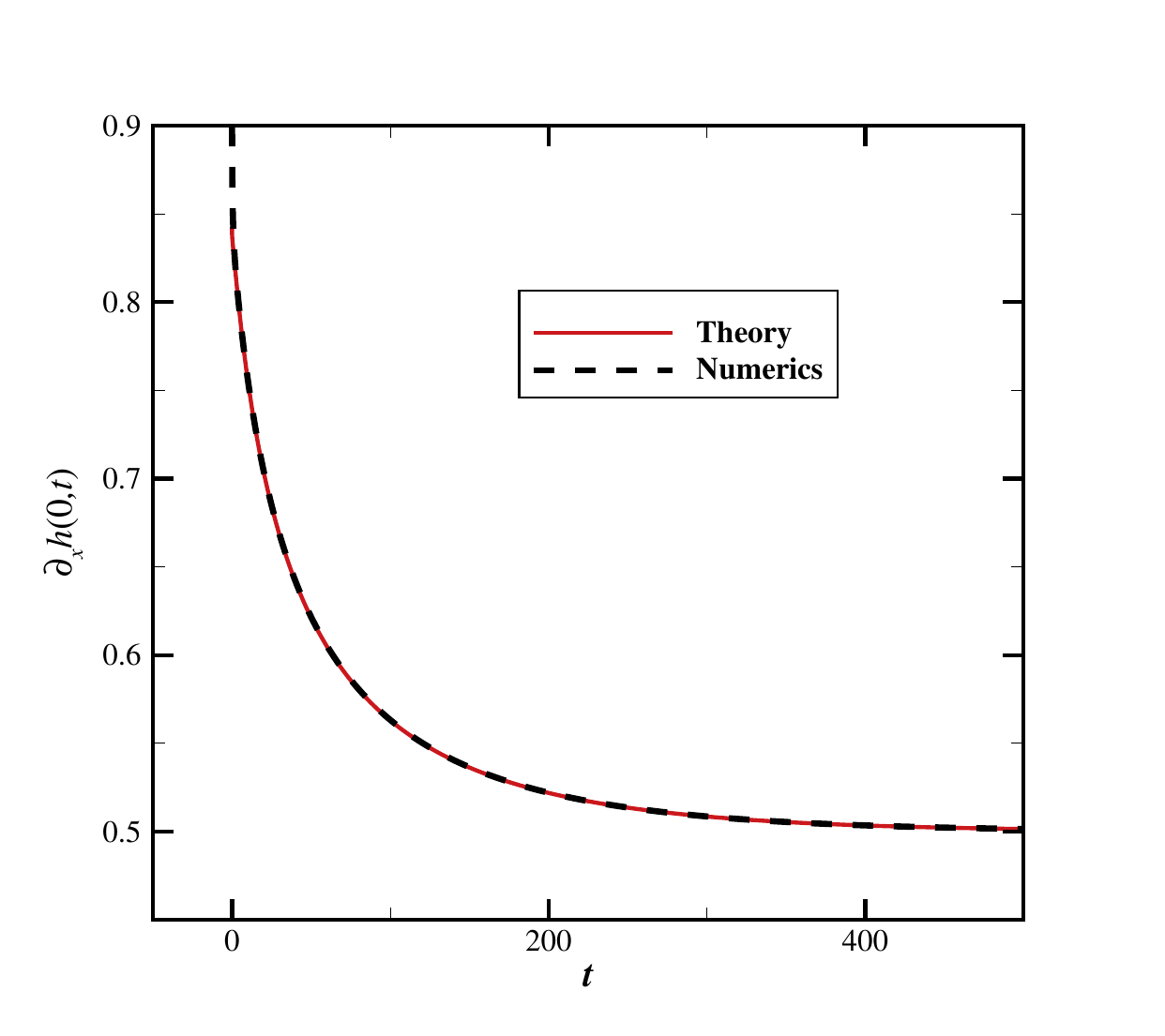}
        \put(-5,87){{\small (\textit{a})}}
    \end{overpic}
    \end{minipage}
    \hfill
    \begin{minipage}{0.45\linewidth}
        \begin{overpic}[width=\linewidth,trim=0.5cm 0.2cm 2cm 1cm,clip]{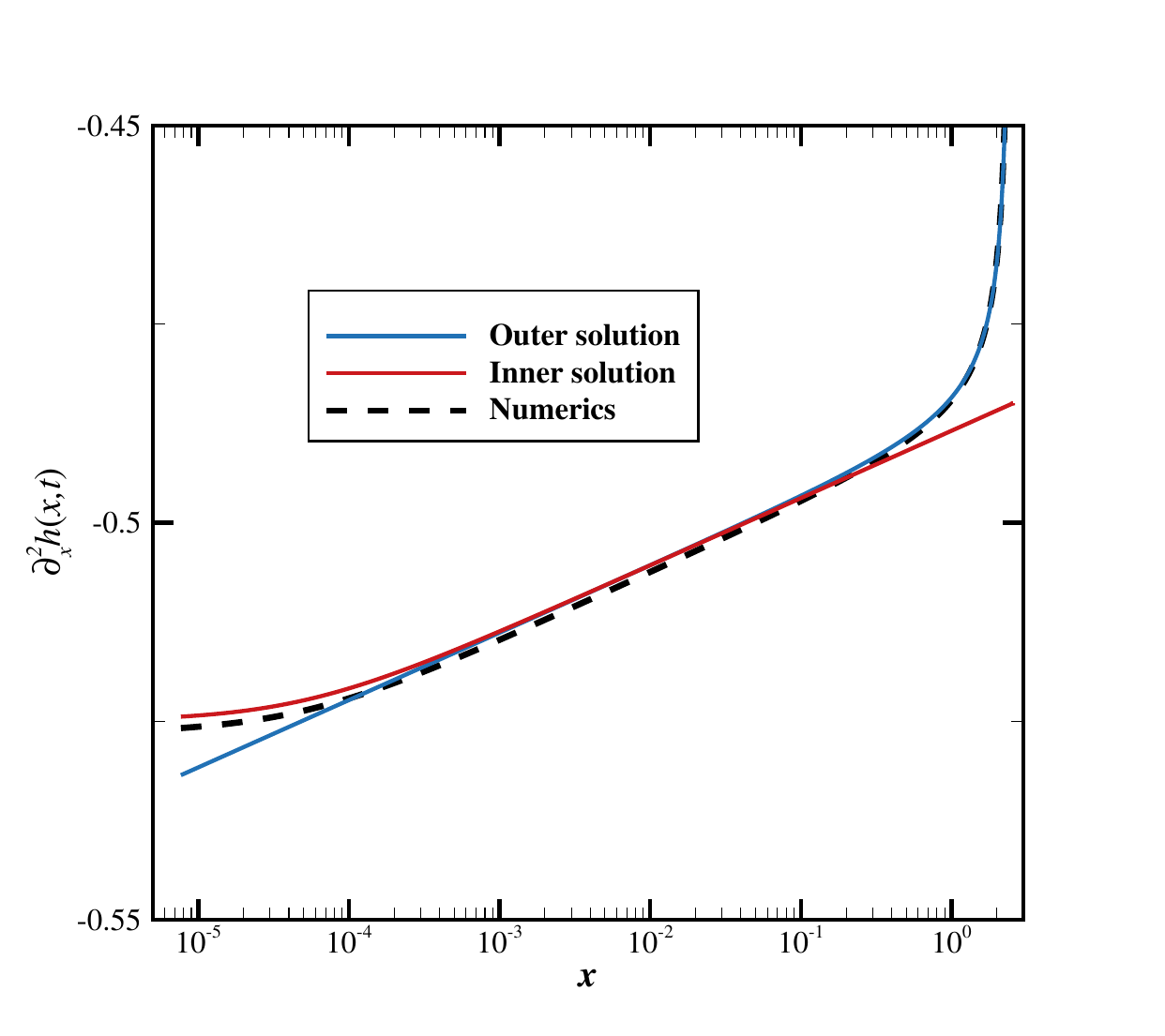}
            \put(-7,87){{\small (\textit{b})}}
        \end{overpic}
    \end{minipage}
    \caption{Microscopic properties of the inner region near the pinned contact line for $K=0.5$. (\textit{a}) Evolution of the contact angle $\partial_xh(0,t)$ with the theoretical curve given by \eqref{thetaTheo}. (\textit{b}) Snapshot of the surface curvature $\partial^2_xh(x,t)$ at $t=50$. The outer solution represents the second-order derivative of \eqref{eq:hpin} together with \eqref{sprh0} and \eqref{sprh1}; the inner solution corresponds to \eqref{eq:curvature}.}
    \label{fig:pinInner}
\end{figure}

\section{3D droplets} \label{Sec:3D}

\begin{figure}
    \centering
    \vspace{1em}
    \begin{minipage}{0.26\textwidth}
        \begin{overpic}[width=\linewidth]{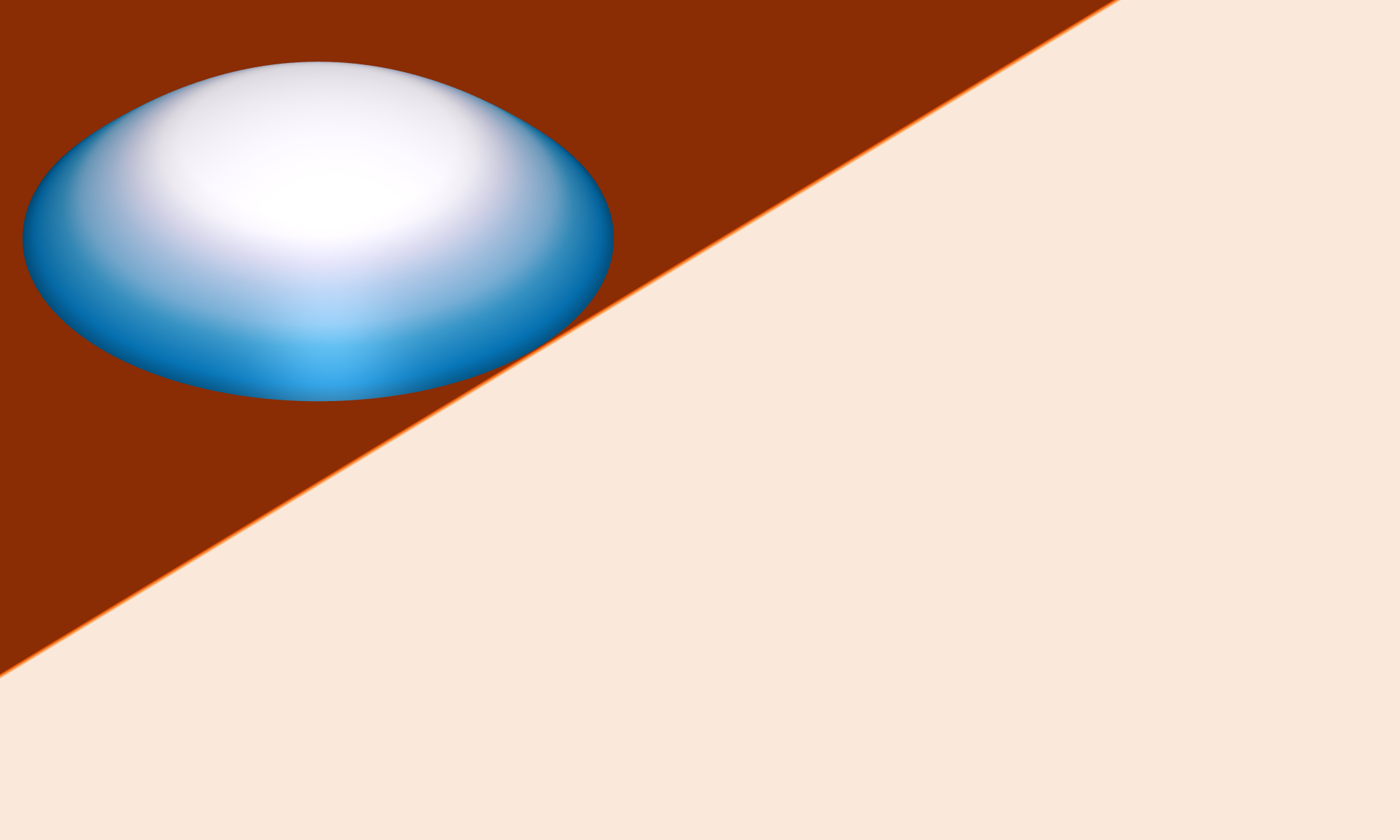}
            \put(-15,53){{\small (\textit{a})}}
            \put(10,5){{\tiny $\boldsymbol{t=0}$}}
        \end{overpic}
    \end{minipage}
    \hspace{.05\textwidth}
    \begin{minipage}{0.26\textwidth}
        \begin{overpic}[width=\linewidth]{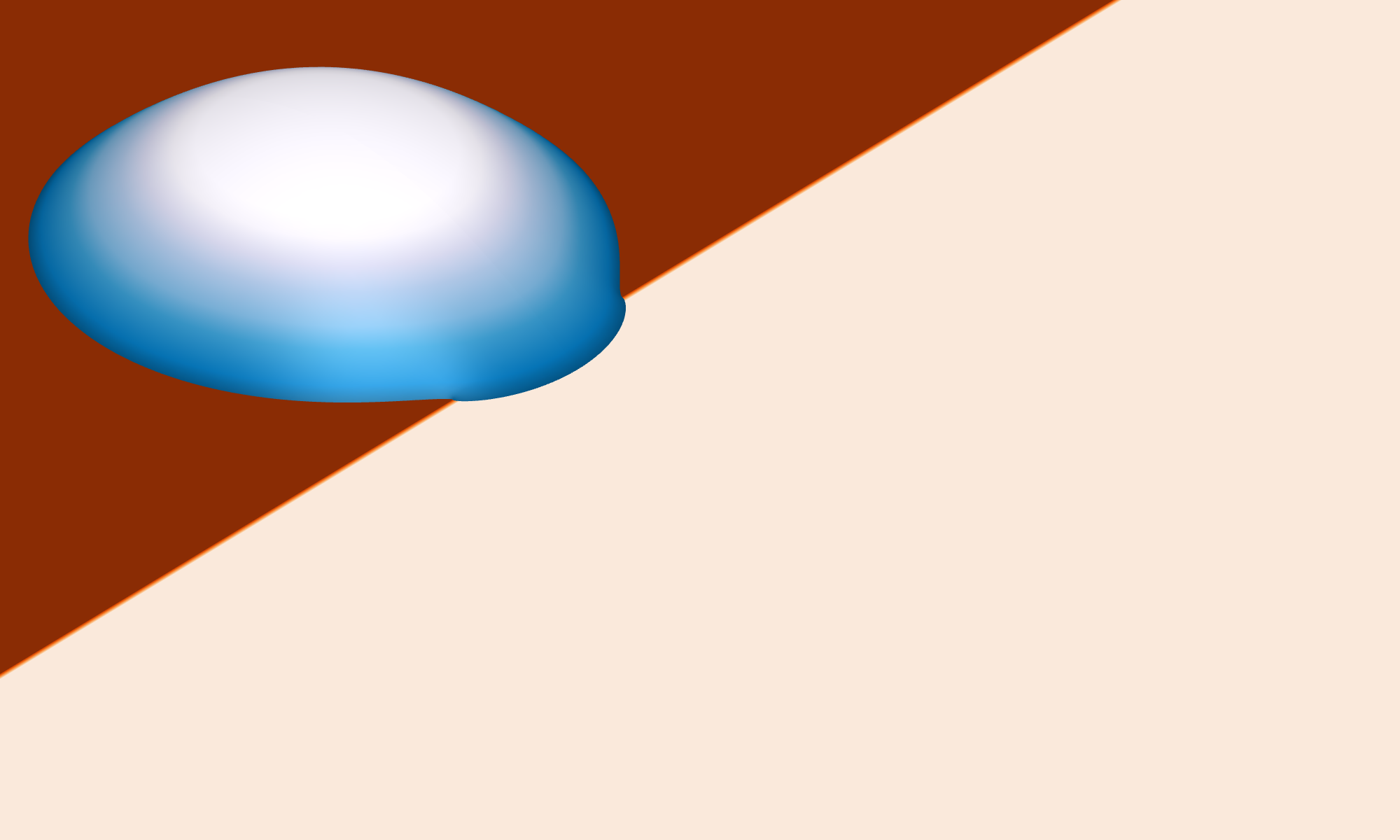}
            \put(-15,53){{\small (\textit{b})}}
            \put(10,5){{\tiny $\boldsymbol{t=10}$}}
        \end{overpic}
    \end{minipage}
    \hspace{.05\textwidth}
    \begin{minipage}{0.26\textwidth}
        \begin{overpic}[width=\linewidth]{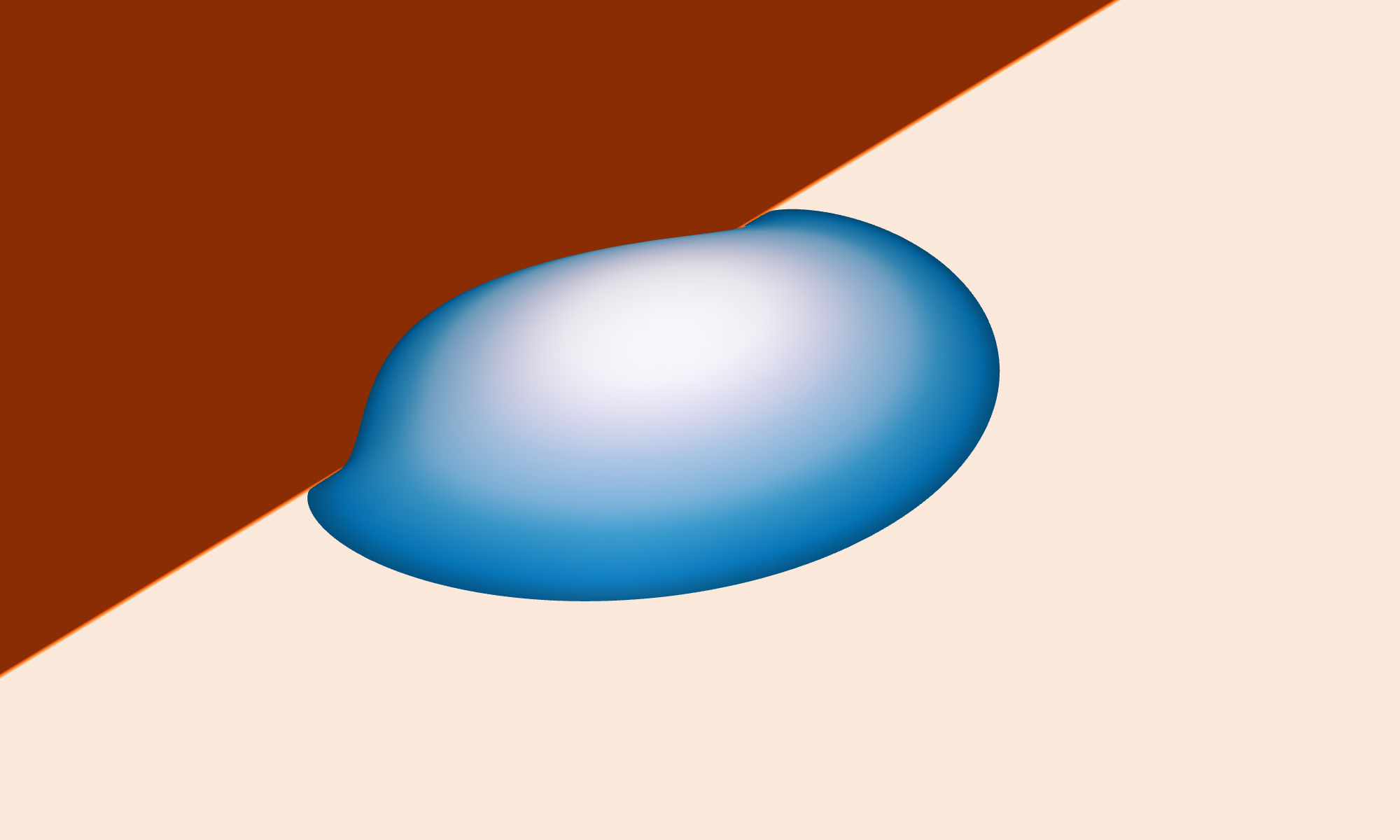}
            \put(-15,53){{\small (\textit{c})}}
            \put(10,5){{\tiny $\boldsymbol{t=100}$}}
        \end{overpic}
    \end{minipage}
    
    \vspace{1em}
    
    \begin{minipage}{0.26\textwidth}
        \begin{overpic}[width=\linewidth]{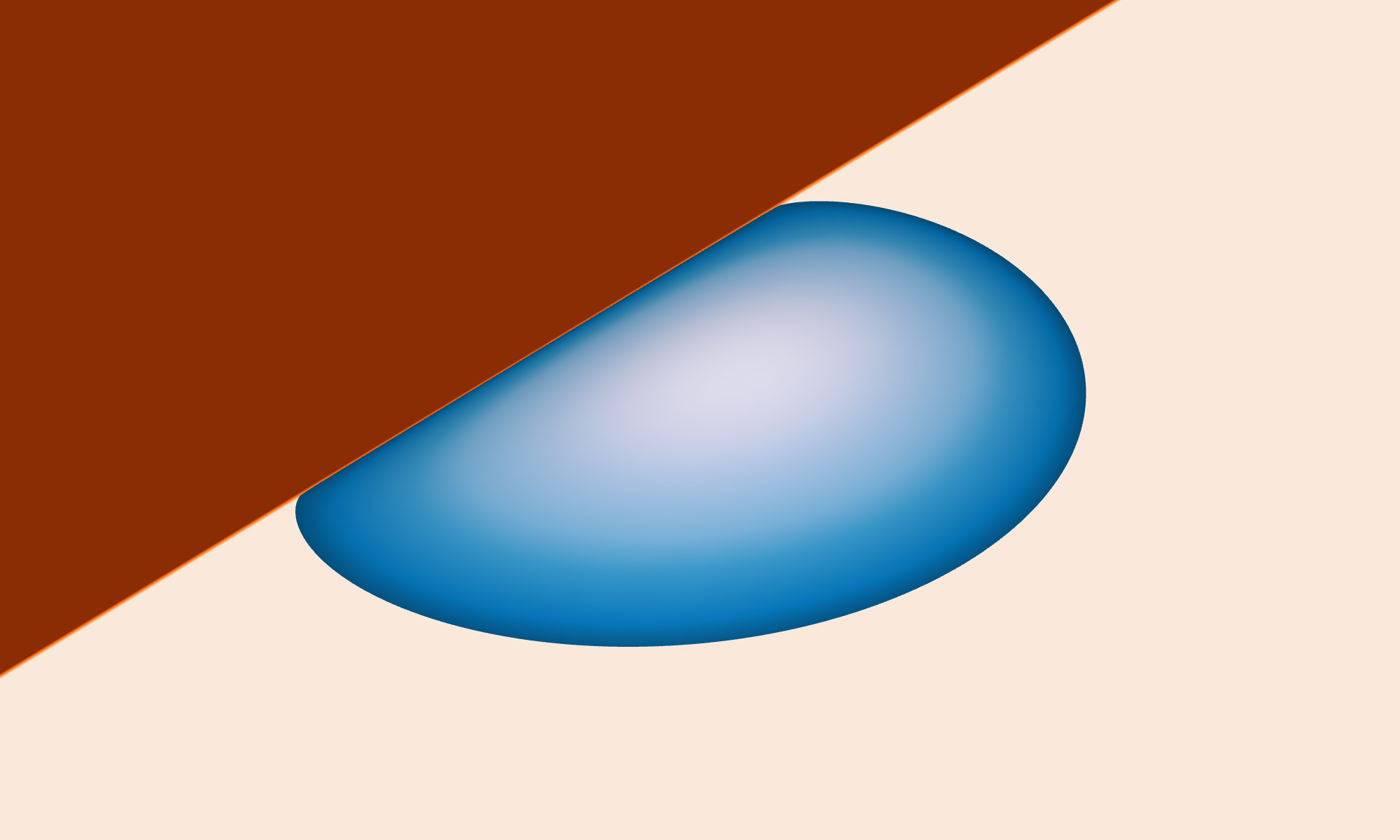}
            \put(-15,53){{\small (\textit{d})}}
            \put(10,5){{\tiny $\boldsymbol{t=130}$}}
        \end{overpic}
    \end{minipage}
    \hspace{.05\textwidth}
    \begin{minipage}{0.26\textwidth}
        \begin{overpic}[width=\linewidth]{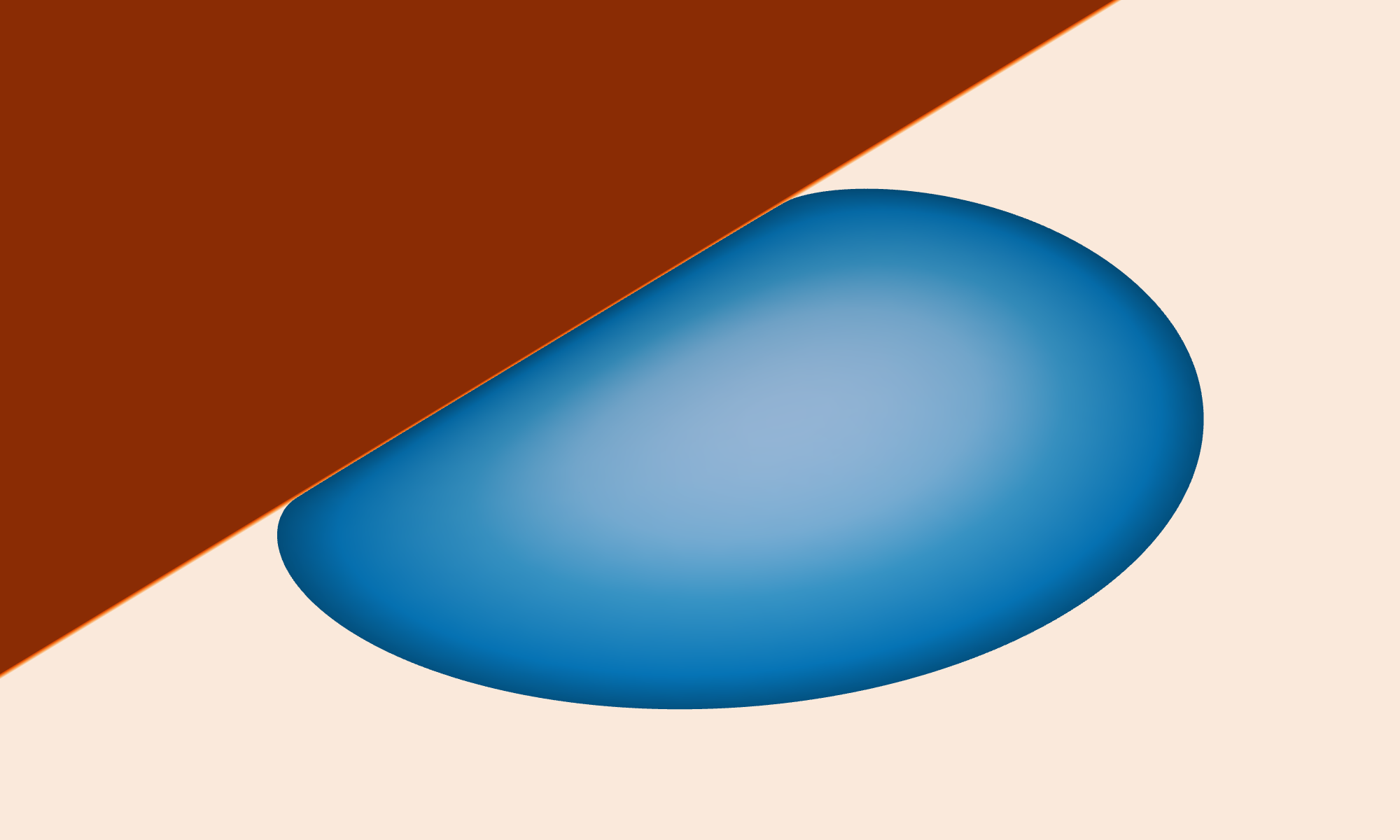}
            \put(-15,53){{\small (\textit{e})}}
            \put(10,5){{\tiny $\boldsymbol{t=300}$}}
        \end{overpic}
    \end{minipage}
    \hspace{.05\textwidth}
    \begin{minipage}{0.26\textwidth}
        \begin{overpic}[width=\linewidth]{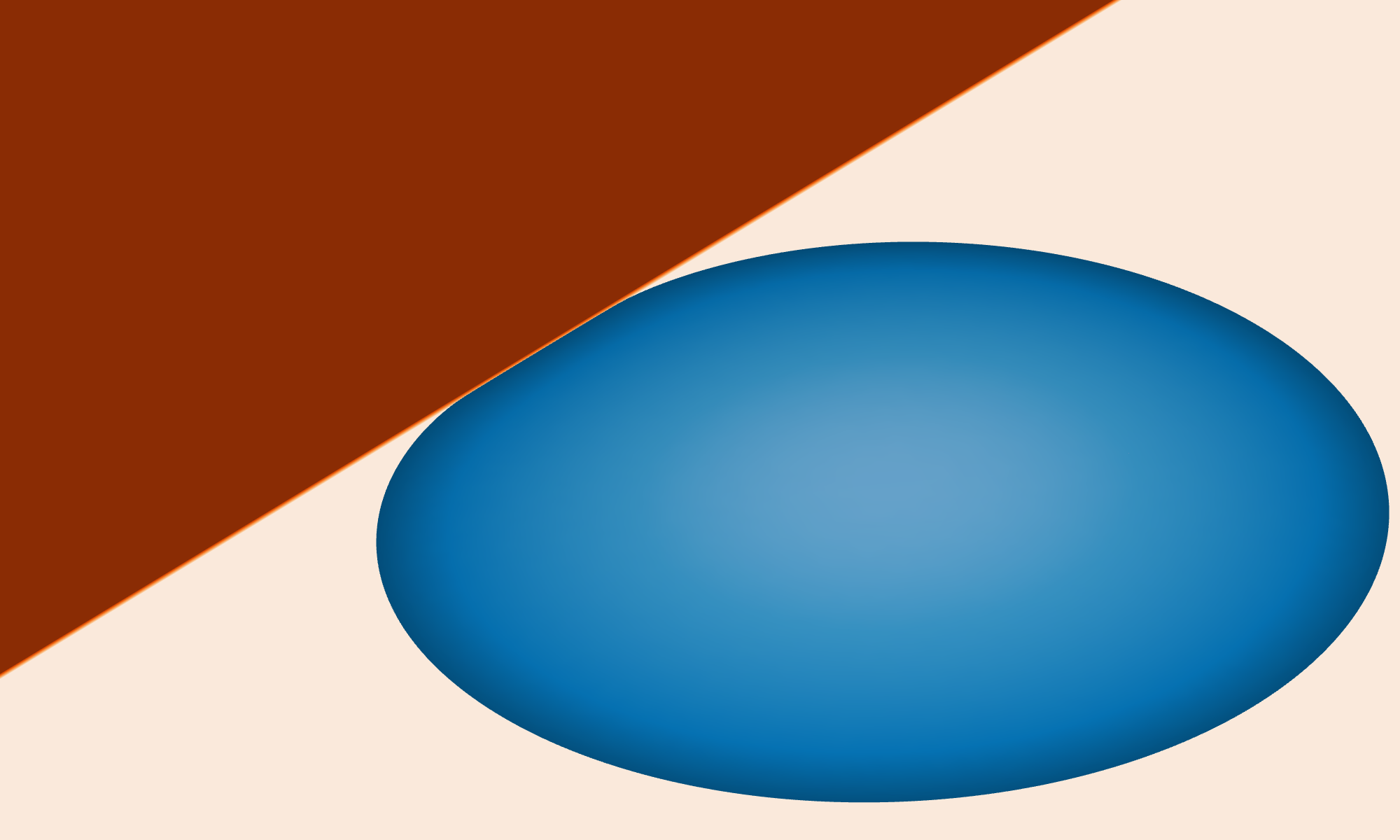}
            \put(-15,53){{\small (\textit{f})}}
            \put(10,5){{\tiny $\boldsymbol{t=9000}$}}
        \end{overpic}
    \end{minipage}
    \caption{Snapshots of a 3D droplet driven by a chemical step with $K=0.2$. The top and bottom row corresponds to the migration and asymmetric spreading stage, respectively.}
    \label{fig:3Dhline}
\end{figure}

For 3D cases, theoretical solutions are difficult to obtain; nonetheless, several quantitative features can be revealed by solving the 3D problem (\ref{3DPDE}) - (\ref{IC}) with the numerical method introduced in \S~\ref{Sec:Eqn}. A symmetric condition has been imposed on the $y=0$ plane to reduce the computational cost. As illustrated in figure~\ref{fig:3Dhline} for $K=0.2$, the two-stage evolution observed in 2D cases is also present for 3D droplets. In the migration stage, the droplet traverses the wettability border. When the droplet completely leaves the less hydrophilic side, the asymmetric spreading stage begins, and the droplet evolves towards equilibrium on the more hydrophilic substrate. Different from spreading on homogeneous substrates, this process is constrained by the wettability border, where part of the contact line is pinned. The droplet eventually detaches from the border at equilibrium.

\begin{figure}
\centering
    \begin{minipage}{0.45\textwidth}
        \begin{overpic}[width=\linewidth,trim=0 9cm 1cm 0,clip]{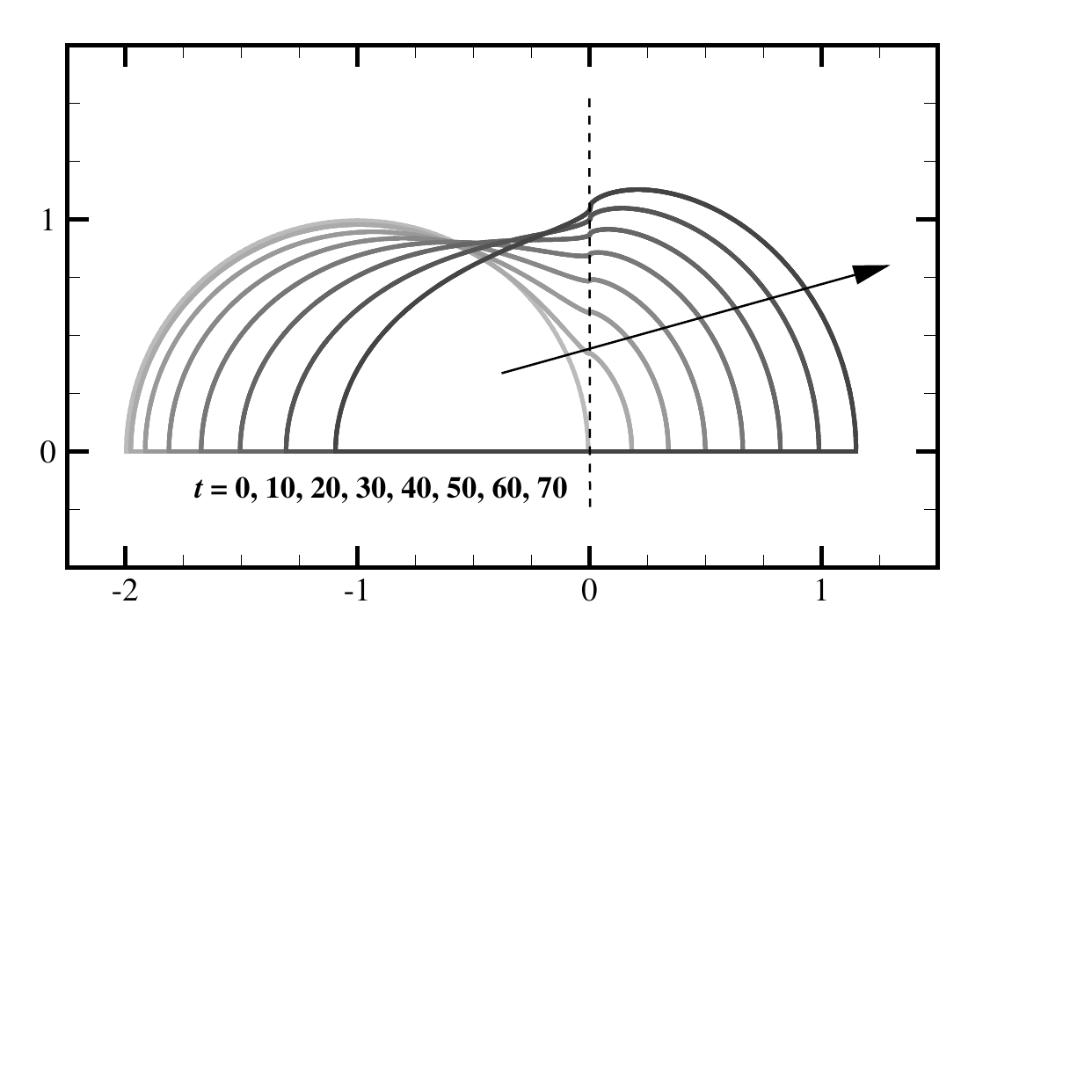}
            \put(-6,53){{\small (\textit{a})}}
        \end{overpic}
    \end{minipage}
    \begin{minipage}{0.45\textwidth}
        \begin{overpic}[width=\linewidth,trim=0 9cm 1cm 0,clip]{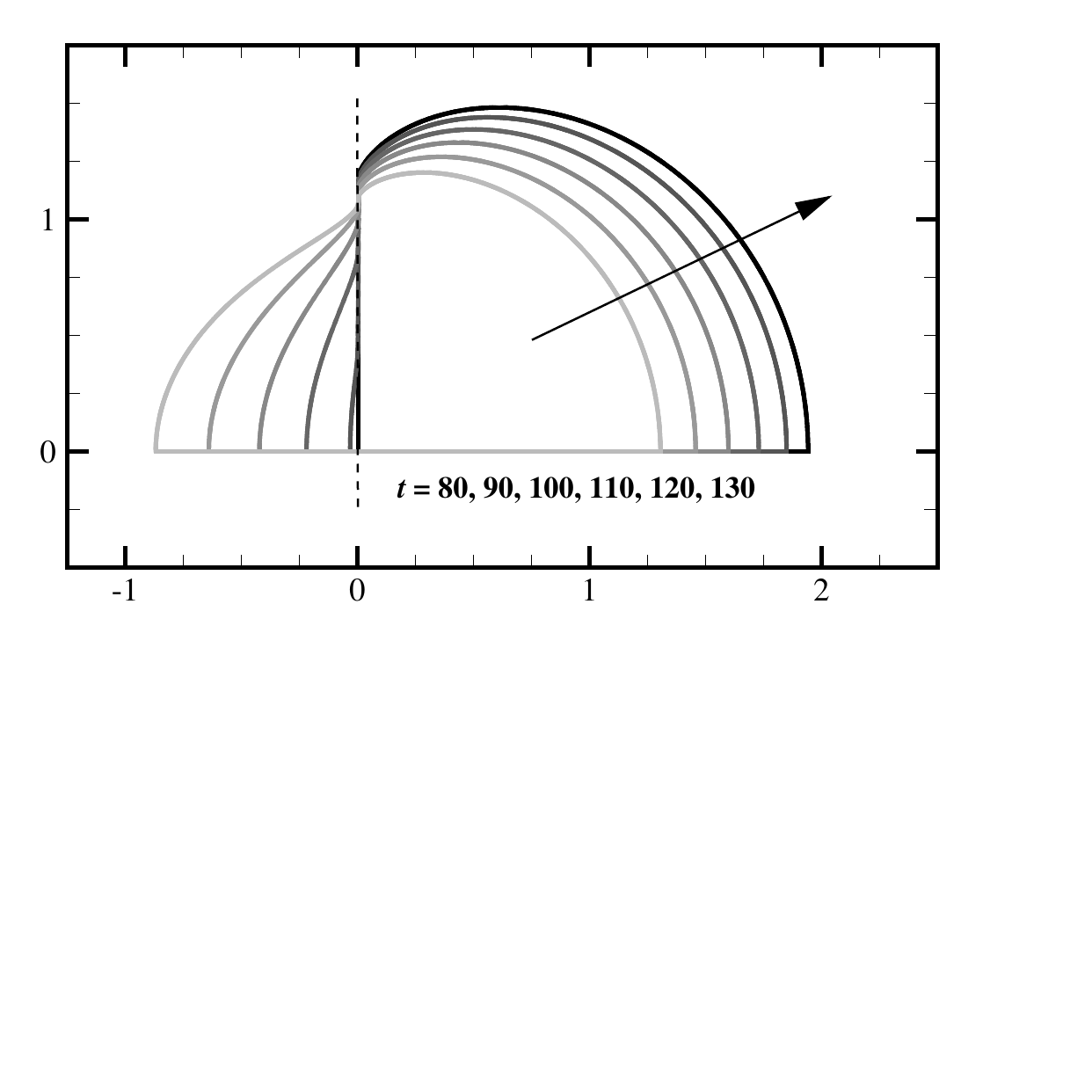}
            \put(-6,53){{\small (\textit{b})}}
        \end{overpic}
    \end{minipage}
    \\
    \begin{minipage}{0.45\textwidth}
        \begin{overpic}[width=\linewidth,trim=0 7cm 1cm 0,clip]{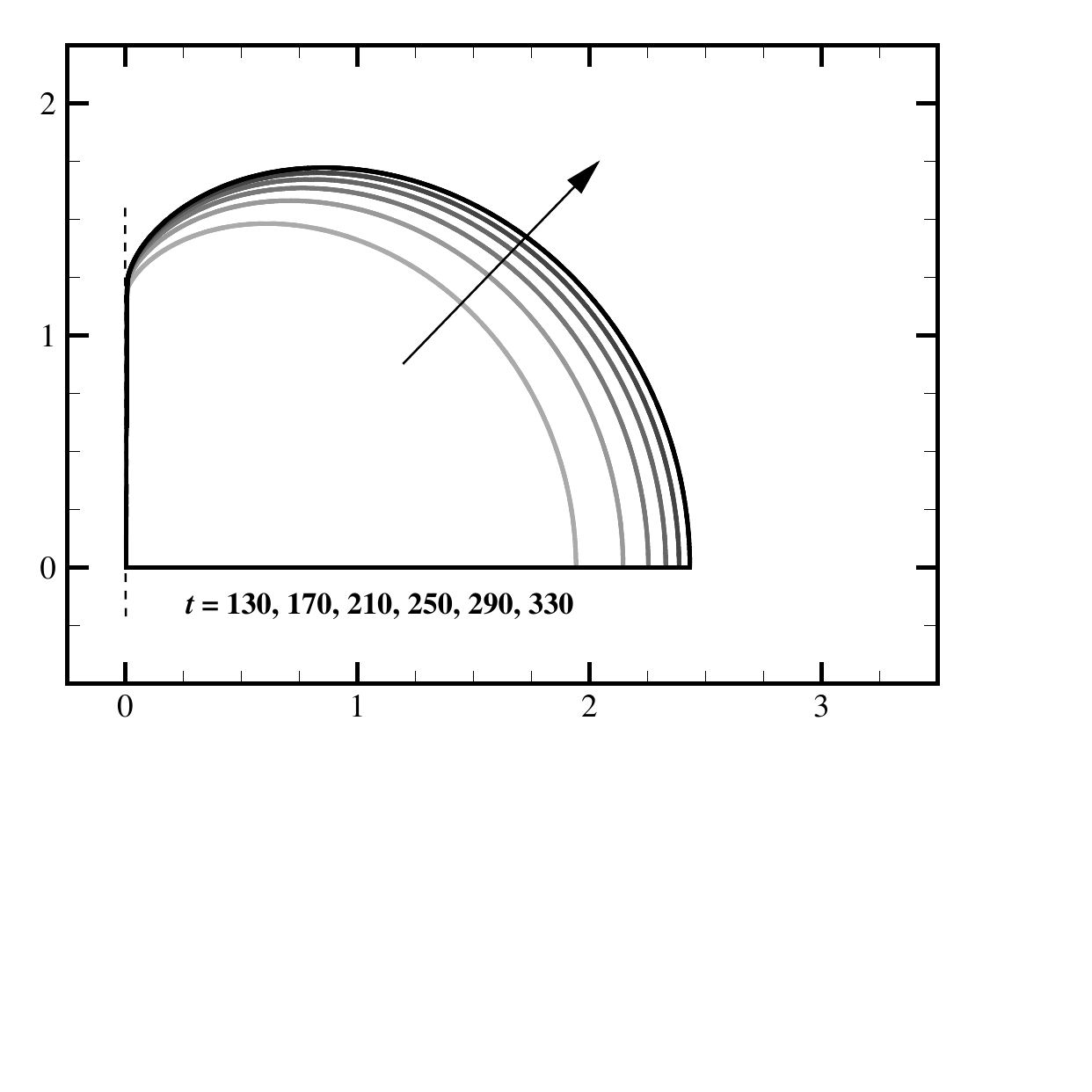}
            \put(-6,62){{\small (\textit{c})}}
        \end{overpic}
    \end{minipage}
    \begin{minipage}{0.45\textwidth}
        \begin{overpic}[width=\linewidth,trim=0 7cm 1cm 0,clip]{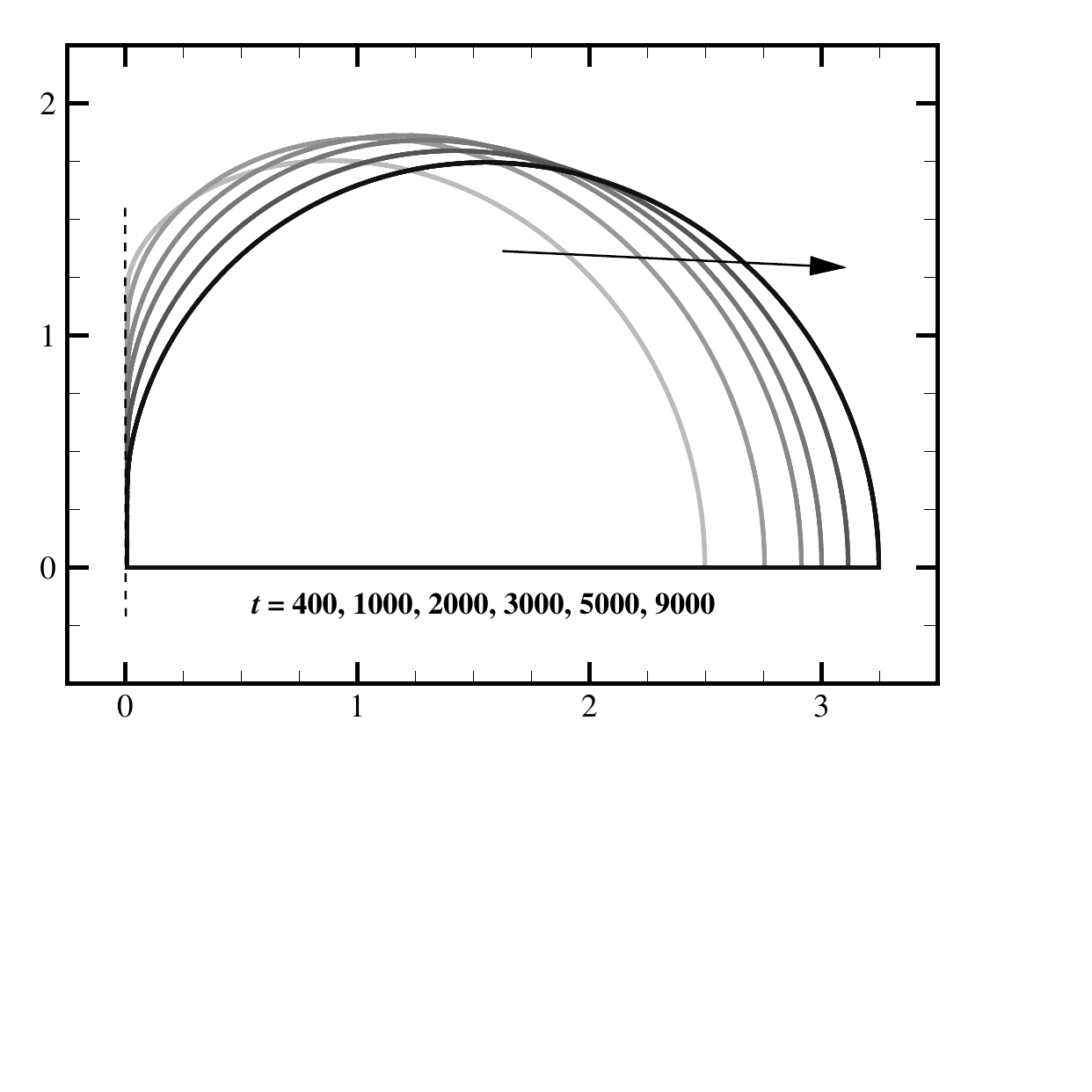}
            \put(-6,62){{\small (\textit{d})}}
        \end{overpic}
    \end{minipage}
  \caption{Snapshots of the contact line  position at different time instants for $K=0.2$.   The arrows indicate the direction of increasing time. Top row: Migration stage, separated by the absence (\textit{a}) and presence (\textit{b}) of pinned contact lines at the wettability border. Bottom row: Asymmetric spreading stage, separated by the absence (\textit{c}) and presence (\textit{d}) of receding contact lines. The dashed lines mark the wettability border.}
\label{fig:CLK2}
\end{figure}

More details of the droplet evolution can be inferred from the morphology of the contact line, as shown in figure~\ref {fig:CLK2}. For clarity, the migration stage is further divided into two sub-stages depending on whether pinned contact lines occur. For $t\lesssim80$ (figure~\ref{fig:CLK2}\textit{a}), the droplet is elongated and moves across the wettability border. The contact line recedes at the droplet tail on the less hydrophilic side, and advances in the more hydrophilic region as well as the less hydrophilic region near the border. At $t\approx80$, only a small fraction of the liquid is left on the less hydrophilic region, where the whole contact line recedes afterwards, while the contact line on the more hydrophilic region keeps advancing. Consequently, the mismatch of the contact line speed causes the occurrence of pinned contact line at the border, as shown in figure~\ref{fig:CLK2}(\textit{b}). The two segments of the pinned contact line extend and eventually connect when the droplet fully enters the more hydrophilic region at $t\approx130$, after which the asymmetric spreading stage starts. The asymmetric spreading is analogously divided into two sub-stages based on the direction of contact line movement. At the beginning of spreading (figure~\ref{fig:CLK2}\textit{c}), the contact line advances, accompanied by a slight increase of the length of the pinned contact line. At later times (figure~\ref{fig:CLK2}\textit{d}), the contact line near the border begins to recede. Eventually, the droplet detaches from the wettability border and approaches the equilibrium state at the more hydrophilic substrate, which corresponds to a circular contact line.

\begin{figure}
    \centering
    \includegraphics[width=.6\linewidth,trim=1cm 0.8cm 1.5cm 1.2cm,clip]{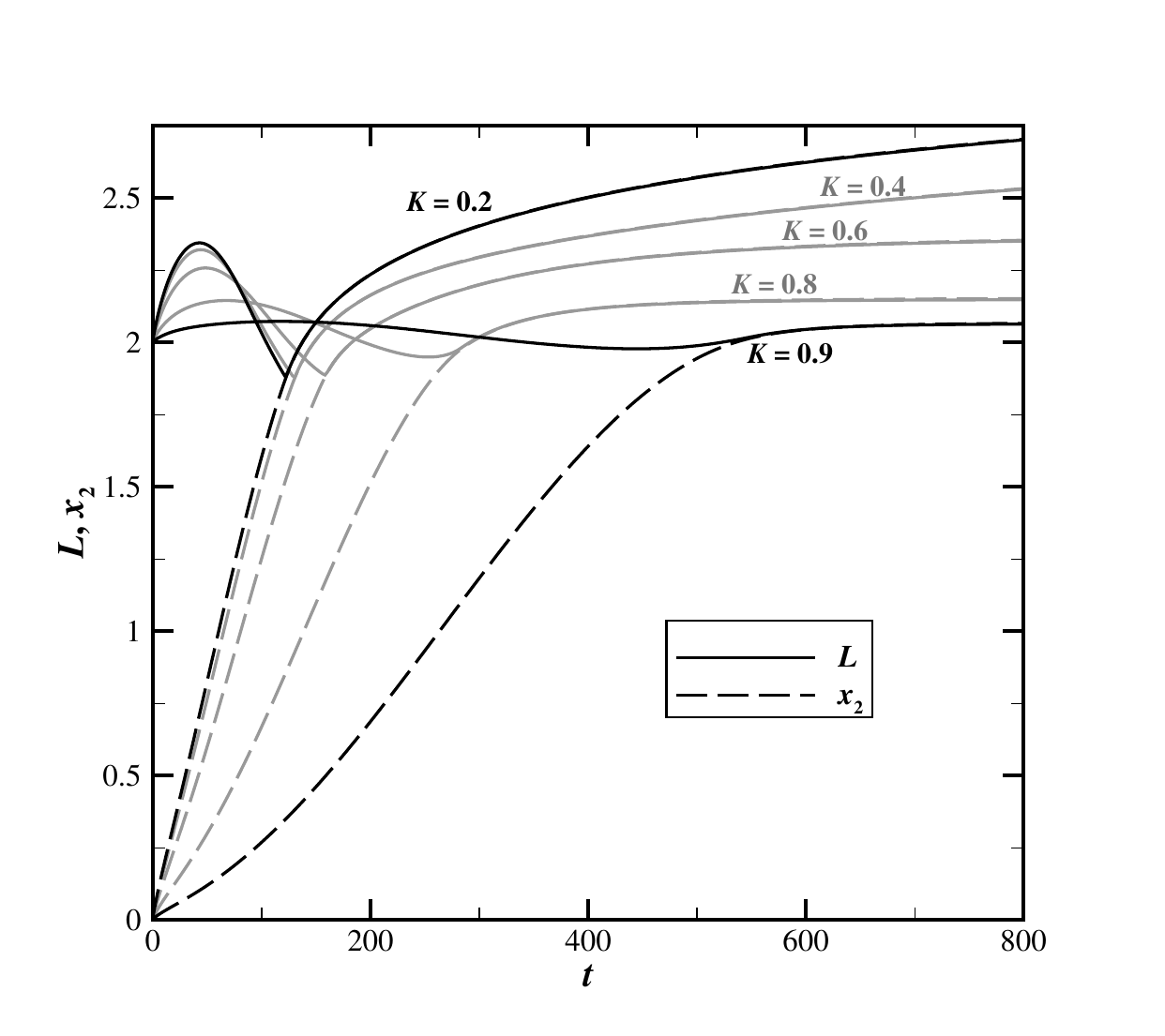}
    \caption{Temporal evolution of droplet length $L$ and the $x$-coordinate of the foremost point $x_2$.}
    \label{fig:3DL_t}
\end{figure}

The temporal evolution of droplet length $L$ and the foremost contact line position $x_2$ is depicted in figure~\ref{fig:3DL_t}. The two curves for each $K$ merge and hence $L=x_2$ in the asymmetric spreading stage. During the migration stage, $x_2$ increases in an approximately linear manner, especially for small values of $K$. The droplet length $L$ exhibits a non-monotonic variation in the migration stage, in contrast to the 2D cases where a steady migration with a constant length is present. A remarkable decrease of $L$ can be observed before the onset of asymmetric spreading stage, leading to a minimum length even less than the initial diameter of the contact line. As can be inferred from figure~\ref{fig:CLK2}(\textit{a}, \textit{b}), this is due to lateral wetting flow on the more hydrophilic region, which increases the longitudinal dewetting speed at the droplet rear. In the asymmetric spreading stage, the droplet length increases towards the equilibrium value.

\begin{figure}
    \centering
    \includegraphics[width=.6\linewidth,trim=1cm 0.8cm 1.2cm 1.2cm,clip]{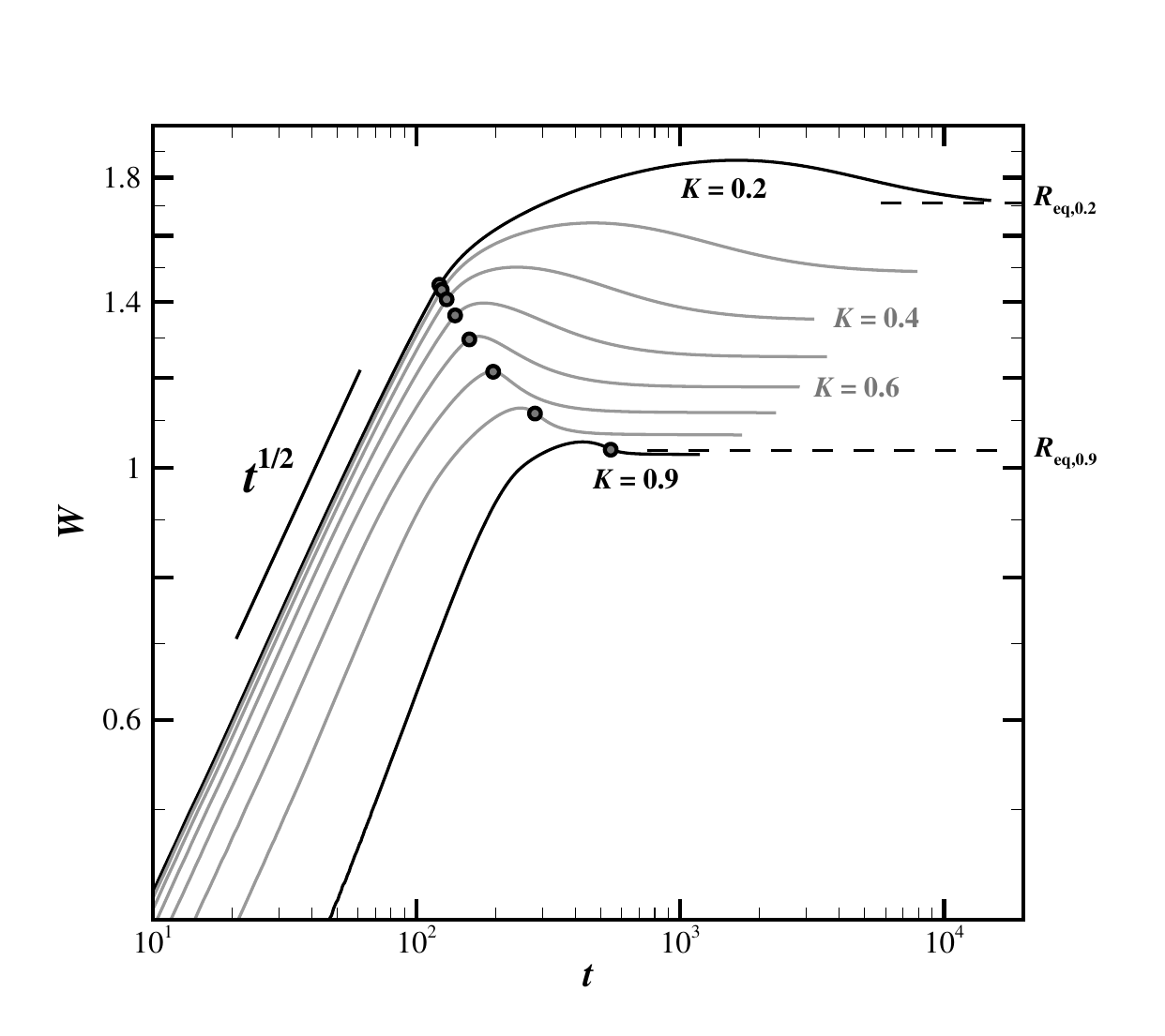}
    \caption{Droplet width on the more hydrophilic side $W$ as a function of time. The grey dots mark the critical instants between two stages. The ratio of contact angles $K$ varies from 0.2 to 0.9 with an increment of 0.1. The equilibrium radius $R_{eq}$ is as marked for $K=0.2$ and $K=0.9$.
    }
    \label{fig:3Dw_t}
\end{figure}

The maximum droplet width on more hydrophilic region $W$ is plotted in figure~\ref{fig:3Dw_t} as a function of $t$ in log--log scales. The grey dots mark the critical instants between the two stages, which correspond to the instants when the leftmost point crosses $x=0$ or equivalently to the merging points of $L$ and $x_2$ in figure~\ref{fig:3DL_t}. At early times, the maximum width occurs at the border and increases following the power law $W\sim t^{1/2}$, reminiscent of the early growth of bridge width in the coalescence of sessile droplets \citep{Ristenpart2006,Kaneelil2022}. The increase of droplet width slows down over time. Interestingly, an overshoot is observed before approaching the equilibrium state. This is again associated with the lateral wetting flow once most of the liquid resides on the more hydrophilic region. 

\section{Effect of initial condition}\label{Sec:IC}

\begin{figure}
    \centering
    \begin{minipage}{0.48\linewidth}
        \centering
        \begin{overpic}[width=\linewidth,trim=1cm 0.8cm 1.5cm 1.2cm, clip]{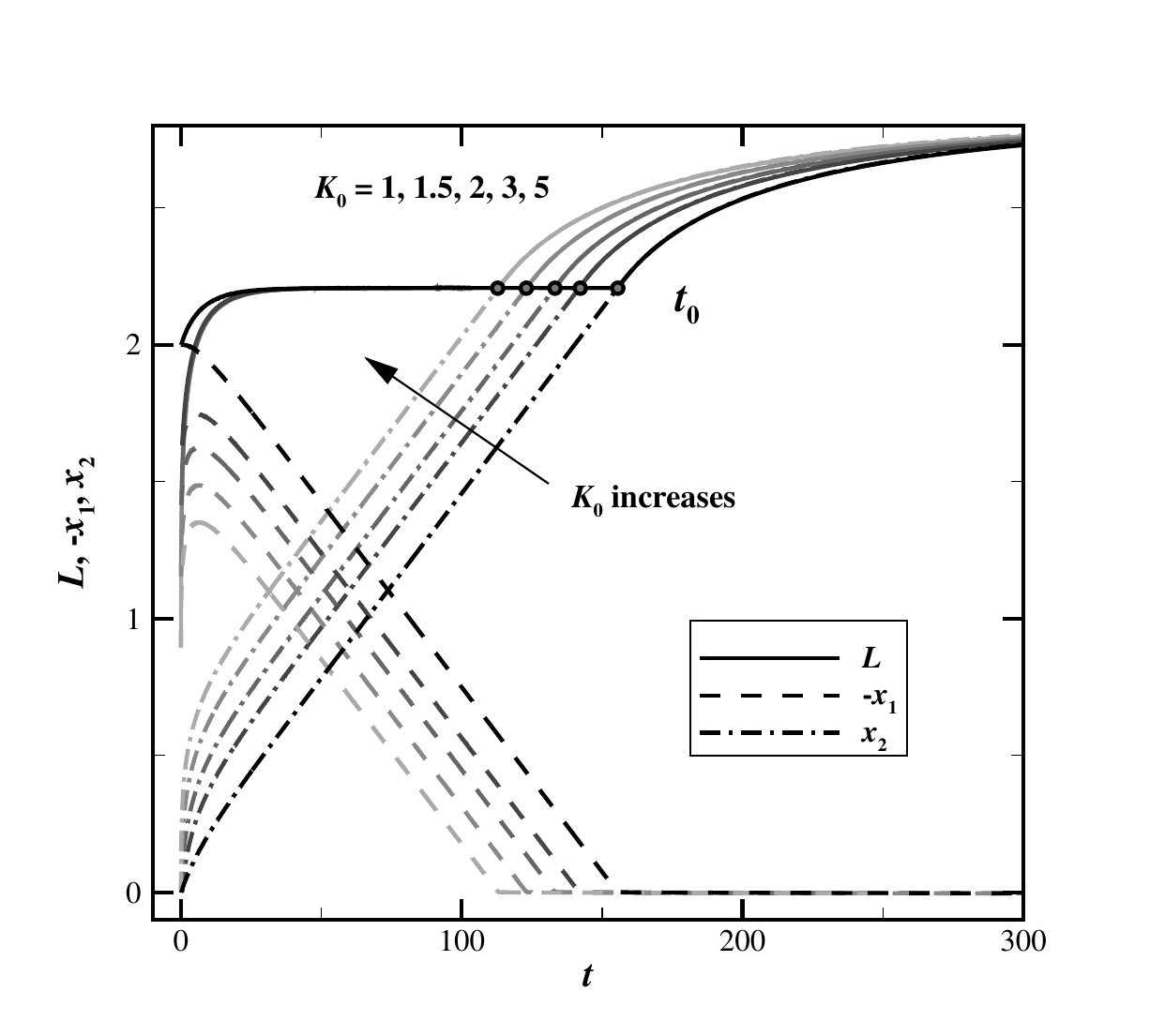}
            \put(-5,81){{\small (\textit{a})}}
        \end{overpic}
    \end{minipage}
    \hfill
    \begin{minipage}{0.48\linewidth}
        \centering
        \begin{overpic}[width=\linewidth,trim=1cm 0.8cm 1.5cm 1.2cm, clip]{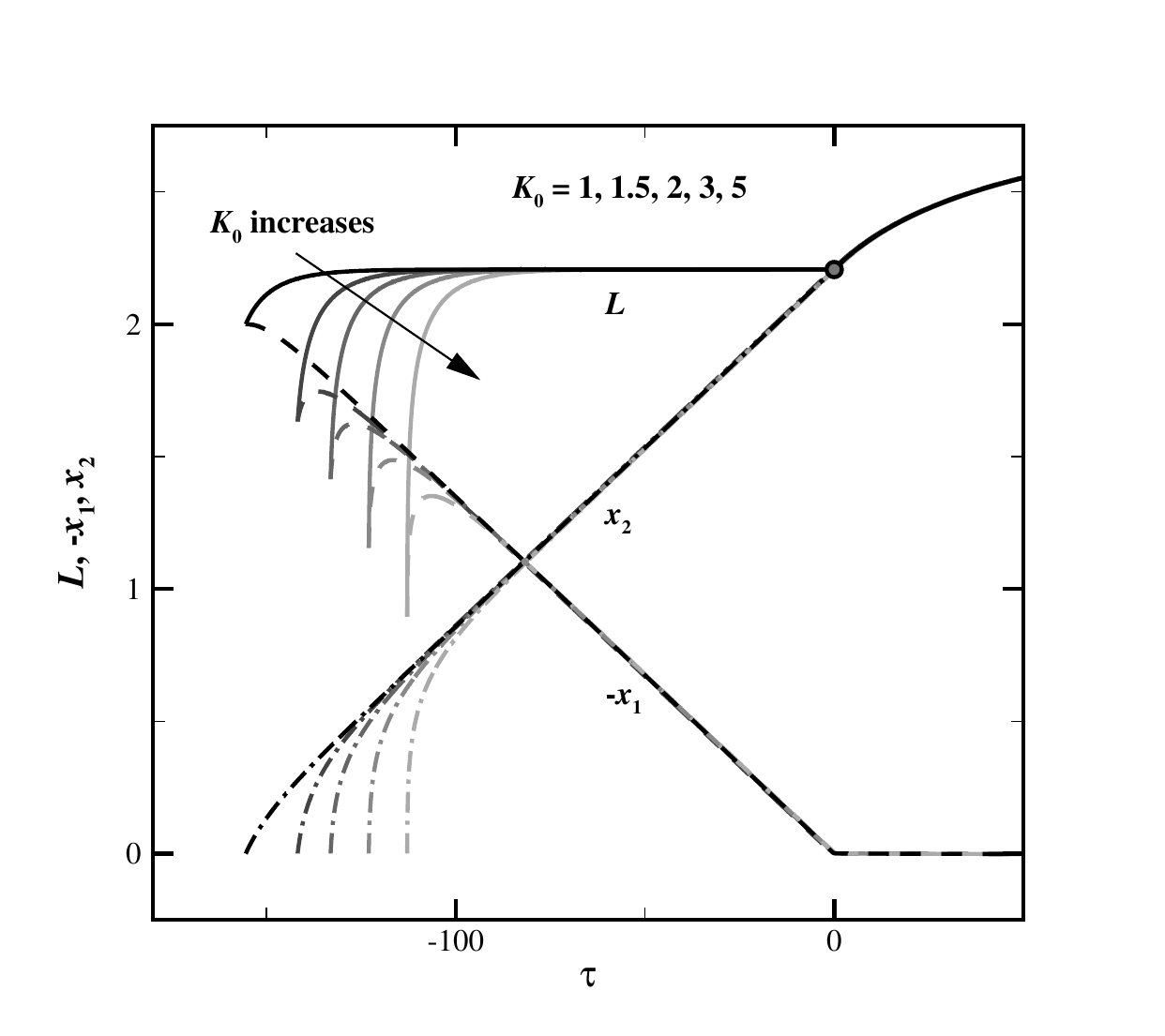}
            \put(-5,81){{\small (\textit{b})}}
        \end{overpic}
    \end{minipage}
    \caption{2D droplet length $L$ and the contact line positions $-x_1$ and $x_2$ as functions of (\textit{a}) time $t$ and (\textit{b}) the shifted time $\tau=t-t_0$ for $K_0\in[1,1.5,2,3,5]$ and $K=0.5$. The grey dots mark the critical instants $t_0$ between the migration and asymmetric stages.}
    \label{fig:Ka2D}
\end{figure}

In this section, the effect of the initial droplet apparent contact angle $K_0$ is investigated by numerically solving the 2D and 3D problems. For both cases, the parameter $K_0$ is varied over the set $[1,1.5,2,3,5]$ and the wettability contrast between two regions is fixed at $K=0.5$. For ease of comparison, a shifted time $\tau=t-t_0$ is introduced, where $t_0$ is the critical instant between the migration stage and the asymmetric spreading stage. Droplet evolution with respect to both $t$ and $\tau$ is plotted and described below.

Figure~\ref{fig:Ka2D} depicts the temporal variation of the contact-line positions $x_1$, $x_2$ and the length $L=x_2-x_1$ of 2D droplets. Regardless of the initial apparent contact angle, the droplet always evolves into a stage of steady migration, followed by a stage of asymmetric spreading. The steady migration and the asymmetric spreading are preceded by an early spreading that elongates the droplet to the same length, which is determined by $K$ as obtained in \S~\ref{Subsec:Mig}. After reaching the same steady migration state, the subsequent droplet motions are identical, as demonstrated in figure~\ref{fig:Ka2D}(\textit{b}) by using the shifted time $\tau$. The differences are only evident in the early spreading process before approaching the steady state. For the case of $K_0=1$, the droplet advances only on the more hydrophilic region, and recedes on the less hydrophilic region. Whereas for $K_0>1$, the droplets initially spread simultaneously on both sides, as indicated by the non-monotonic variation of $x_1$. As $K_0$ increases, the droplet is more plump upon contacting with the wettability border, hence both contact lines advance more rapidly and the elongation accelerates, causing a faster early spreading. The transient advancing of the left contact line is reversed by the overwhelming effect of the wettability contrast and the steady migration develops. As the droplet at larger $K_0$ is narrower, it leaves the less hydrophilic region at an earlier time $t_0$.

\begin{figure}
    \centering
    \begin{minipage}{0.48\linewidth}
        \centering
        \begin{overpic}[width=\linewidth,trim=1cm 0.8cm 1.5cm 1.2cm, clip]{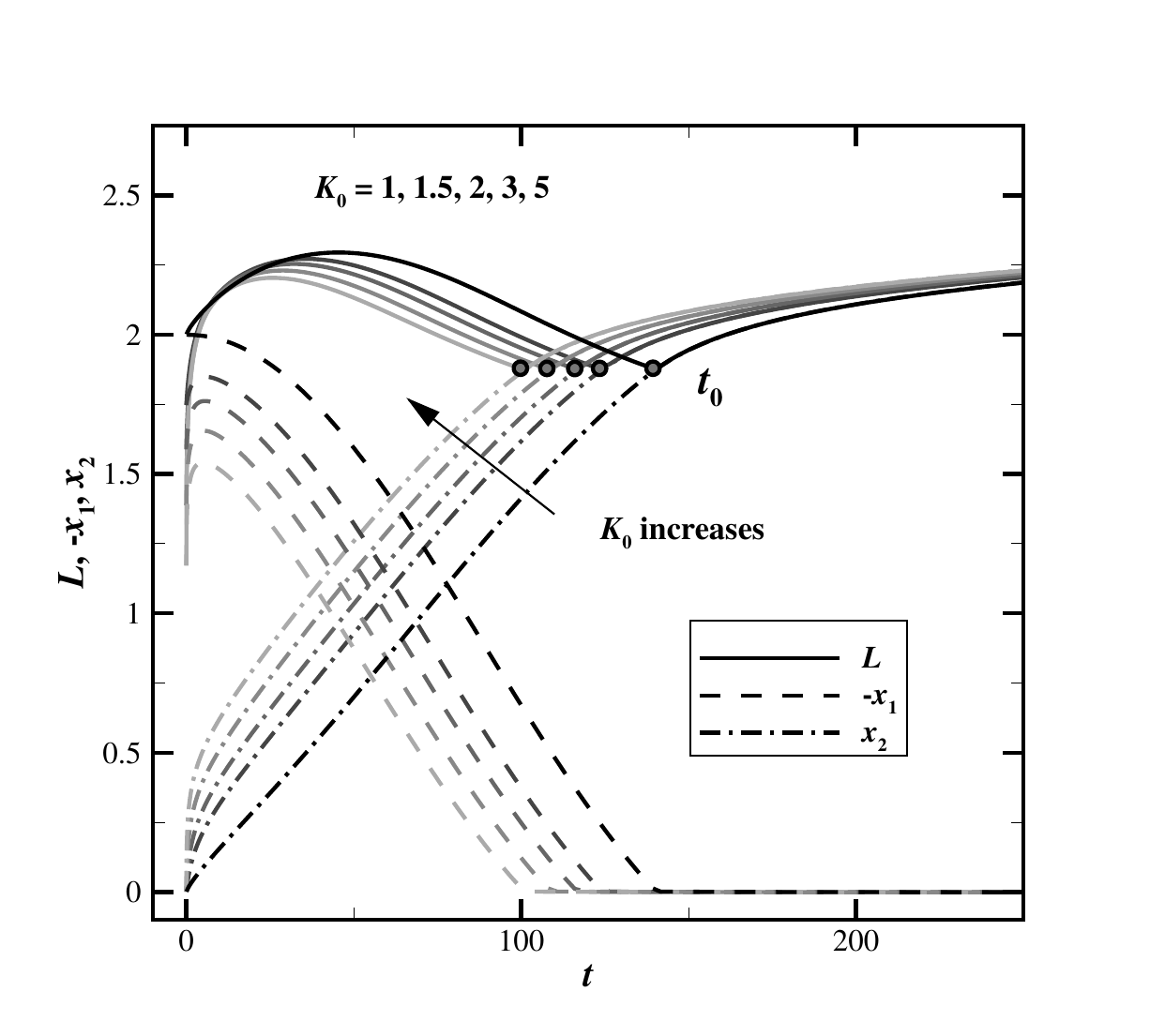}
            \put(-5,81){{\small (\textit{a})}}
        \end{overpic}
    \end{minipage}
    \hfill
    \begin{minipage}{0.48\linewidth}
        \centering
        \begin{overpic}[width=\linewidth,trim=1cm 0.8cm 1.5cm 1.2cm, clip]{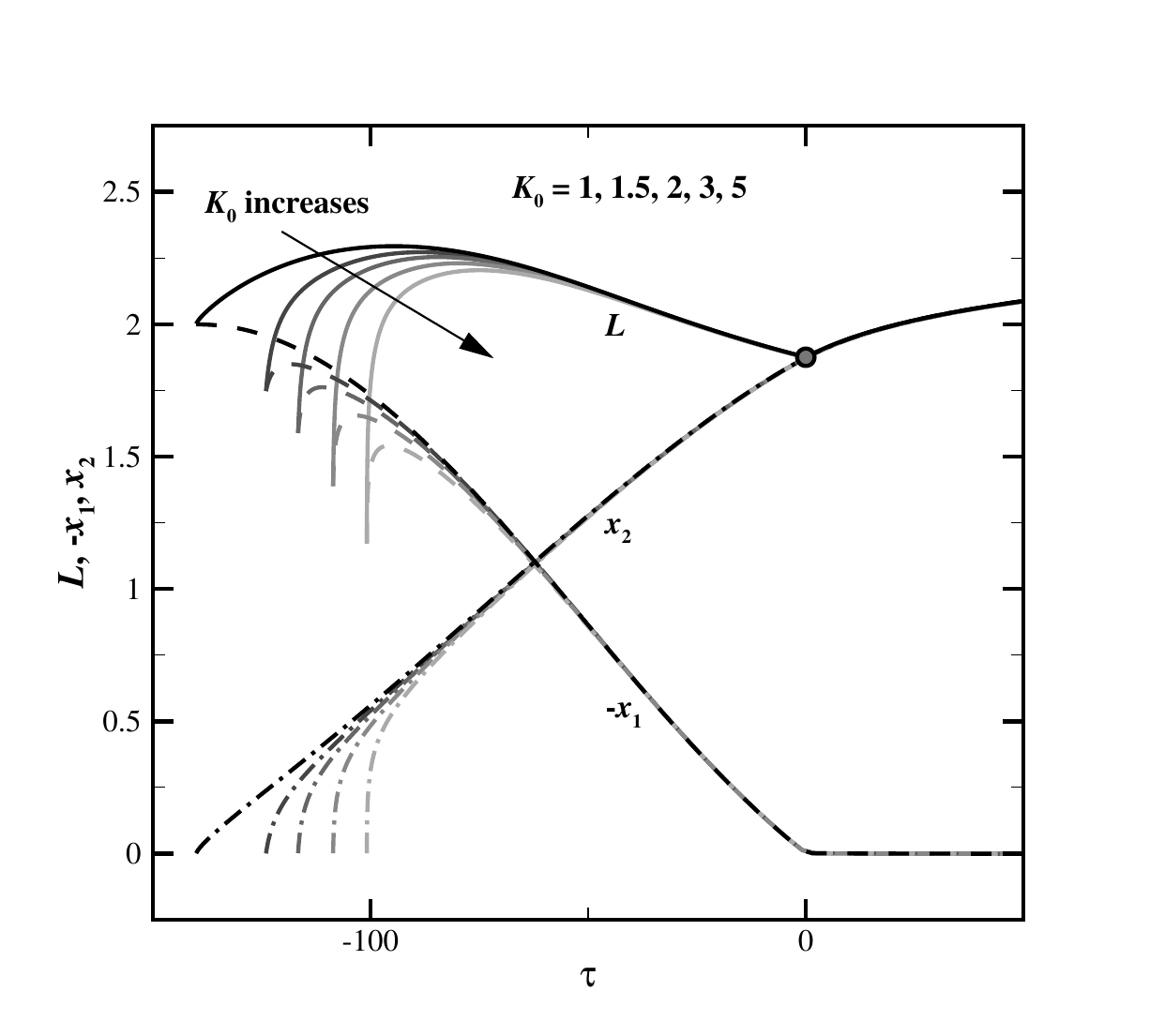}
            \put(-5,81){{\small (\textit{b})}}
        \end{overpic}
    \end{minipage}
    \\
    \begin{minipage}{0.48\linewidth}
        \centering
        \begin{overpic}[width=\linewidth,trim=1cm 0.8cm 1.5cm 1.2cm, clip]{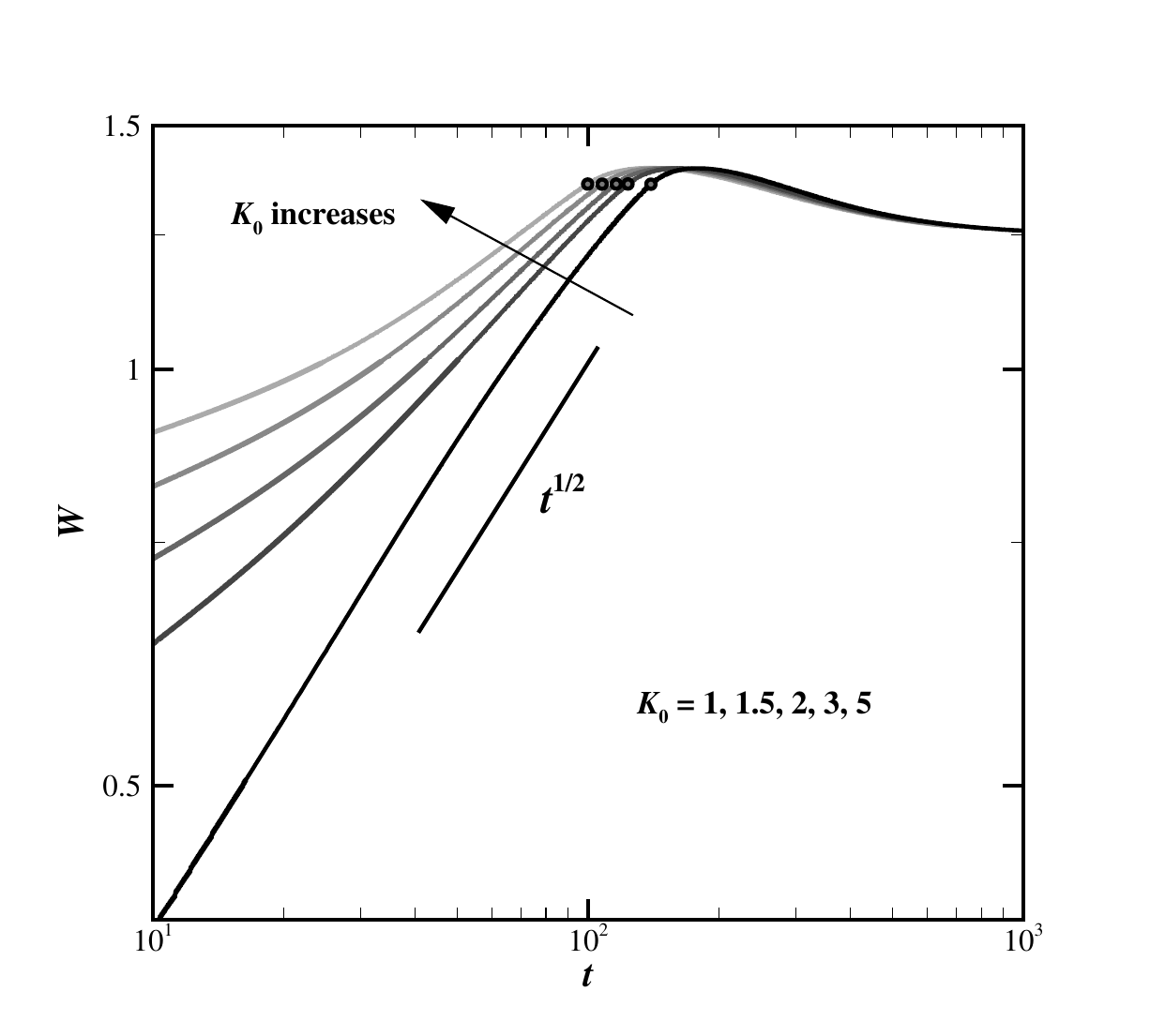}
            \put(-5,81){{\small (\textit{c})}}
        \end{overpic}
    \end{minipage}
    \hfill
    \begin{minipage}{0.48\linewidth}
        \centering
        \begin{overpic}[width=\linewidth,trim=1cm 0.8cm 1.5cm 1.2cm, clip]{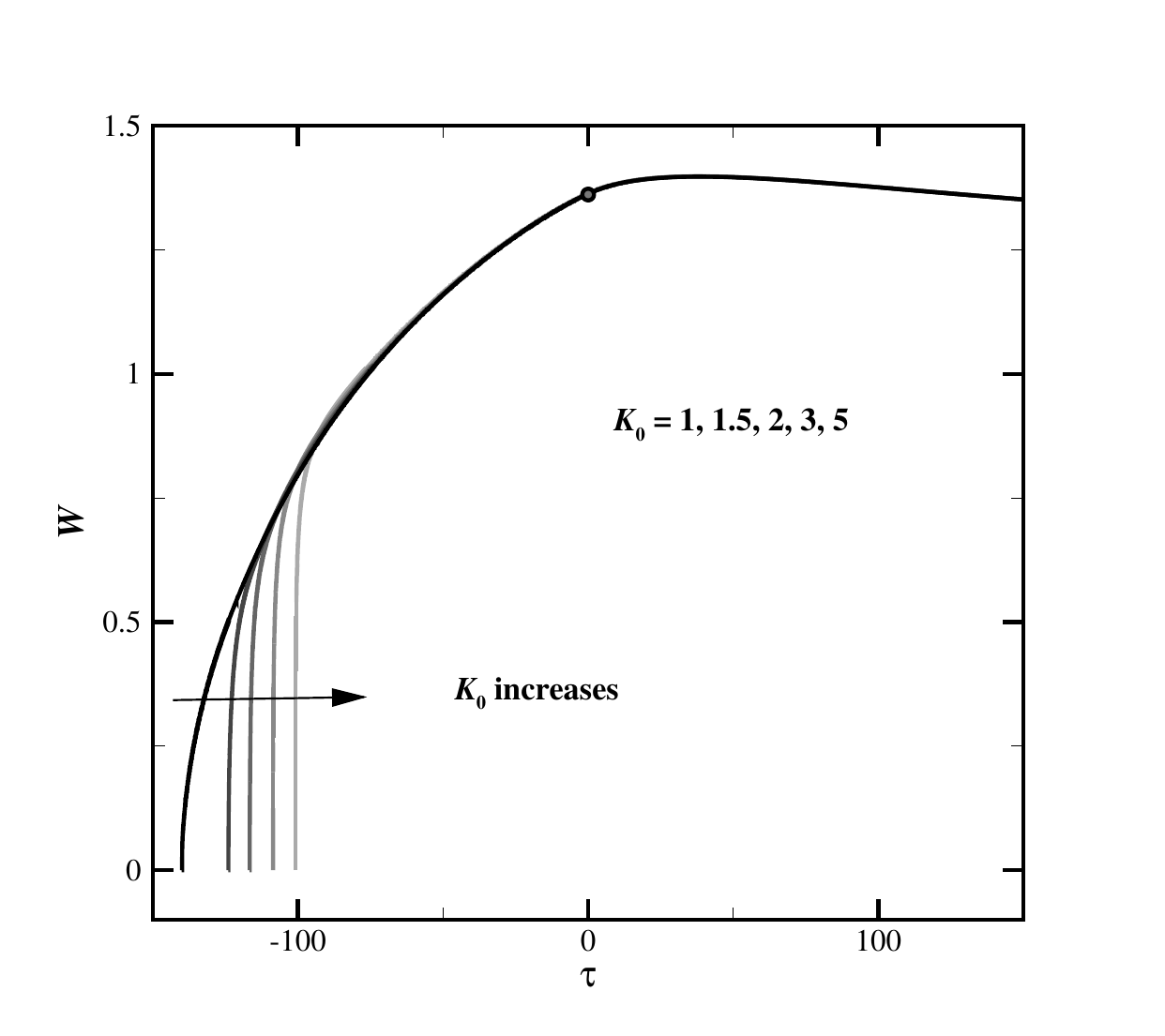}
            \put(-5,81){{\small (\textit{d})}}
        \end{overpic}
    \end{minipage}
    \caption{3D droplet size and position as functions of time $t$ (the left column) and the shifted time $\tau$ (the right column) under different $K_0\in[1,1.5,2,3,5]$ and fixed $K=0.5$. (\textit{a, b}) Droplet length $L$ and the $x$-coordinates of the backmost point $x_1$ and the foremost point $x_2$. (\textit{c}) Droplet width $W$ in log-log coordinates. (\textit{d}) Droplet width $W$ as a function of shifted time $\tau$ in linear coordinates. The grey dots mark the critical instants $t_0$ between the migration and asymmetric stages.}
    \label{fig:Ka3D}
\end{figure}

For 3D droplets shown in figure~\ref{fig:Ka3D}, the effect of $K_0$ is also confined to the early process. Figure~\ref{fig:Ka3D}(\textit{a}, \textit{b}) demonstrate the droplet evolution along the $x$-direction, where the length varies non-monotonically during migration, but qualitatively similar $K_0$-dependence to the 2D cases are observed. For different $K_0$, the droplet length and $x$-coordinates still converge at some point and follow the same dynamics thereafter when the shifted time is used. The droplet width $W$ defined in the $y$-direction, shown in figure~\ref{fig:Ka3D}(\textit{c}, \textit{d}), grows more rapidly for larger $K_0$ in the early spreading process, but eventually converges during the migration stage before $\tau=0$. When $K_0>1$, the $t^{1/2}$-law of the droplet width $W$ is violated at early times, presumably owing to the significant spreading on both regions.

As demonstrated above, the value of $K_0$ mainly influences the early spreading process that adjusts the droplet profile to reach the migration state, after which the evolution is determined by $K$ rather than $K_0$. 

\section{Concluding remarks} \label{Sec:conclu}
We have investigated the spontaneous relaxation of a thin droplet over a chemical step and identified two distinct stages for both 2D and 3D configurations. The contact line singularity is alleviated using a Navier slip model with a sufficiently small slip length. Driven by the wettability contrast, the droplet first migrates from the less to the more hydrophilic side. Under the same wettability contrast $K$, droplets with different initial shapes characterized by varying $K_0$ reaches the identical state and follow the same migration motion thereafter. When all the liquid leaves the less hydrophilic region, the droplet remains far from equilibrium, leading to a subsequent stage of asymmetric spreading with a portion of contact line pinned at the border.

For 2D droplets, a steady motion is reached during the migration stage, where the droplet maintains a stationary profile that travels at a constant velocity. As demonstrated by asymptotic analysis and numerical calculations, both the droplet length and velocity increase with decreasing $K$, until attaining their maximum values when the more hydrophilic region is perfectly wettable. During the 2D asymmetric spreading, the droplet evolution is primarily determined by the advancing contact line, while the detailed information near the pinned contact line is not involved. Our asymptotic analysis reveals the existence of a boundary layer near the pinned contact line, across which the surface slope is approximately constant, whereas the curvature exhibits logarithmic variations and can be regularized by introducing the Navier slip condition.

The evolution of 3D droplets is qualitatively analogous to the 2D case, while being significantly affected by the lateral flow. When $K_0=1$, the contact line of the droplet on the more hydrophilic region advances linearly in  the $x$-direction, and expands according to a power law $t^{1/2}$ in  the $y$-direction at early times. Different from 2D droplets, the lateral wetting flow in the more hydrophilic region causes a non-monotonic variation of the droplet length and complicated formation of the pinned contact line. When most of the liquid has reached the more hydrophilic region, the droplet length decreases and a portion of contact line is pinned at the wettability border. The presence of the pinned contact line results in the asymmetric spreading of the droplet, which eventually detaches from the border and approaches a new equilibrium state. An overshoot of the droplet width occurs due to the lateral wetting flow. 

We note that 3D perfect wetting case ($K=0$) is not investigated in the present work owing to the limitations of the numerical method. Nonetheless, qualitative behaviour can be deduced based on the results of partially wetting. Since the contact line must advance indefinitely on perfectly wetting substrate, the detaching process observed in figure~\ref{fig:CLK2}(\textit{d}) will be absent. Instead, the process in figure~\ref{fig:CLK2}(\textit{c}) is expected to sustain, and eventually the liquid spreads and covers the more hydrophilic half plane completely. In this scenario, the width of the pinned contact line at the wettability border and the evolution of the film morphology are worth investigating. 

Finally, it is noteworthy that the expression of the boundary layer near the pinned contact line \eqref{eq:curvature} can be generalized to general macroscopic configurations. We expect that the surface profile near the pinned contact lines depends primarily on the local behaviours of the macroscopic surface, e.g., the apparent contact angle $\theta(t)$, which also represents the slope angle across the boundary layer, and the apparent curvature $\kappa_{ap}(t)$. Using these local quantities, we can write in dimensional form
\begin{equation}
    \partial^2_xh=\kappa_{ap}-\frac{3\mu\dot{\theta}}{2\sigma\theta^3}\ln\frac{3\lambda/\theta+x}{PL_M},
\end{equation}
where $L_M$ is a macroscopic length scale and the coefficient $P$ depends on the specific macroscopic configuration. This relation holds for slow rotational speed $\dot{\theta}$.

\begin{bmhead}[Funding.]
This work was supported by the NSFC (grant nos 12241204, 12325208 and 12388101).
\end{bmhead}

\begin{bmhead}[Declaration of interests.]
The authors report no conflict of interest.
\end{bmhead}

\begin{appen}

\section{Numerics for 2D migration}\label{appC}
This section introduces the numerical method for solving (\ref{migODE}-\ref{volCons}). As $L$ is an unknown variable, the domain $x\in[0,L]$ was rescaled by $r=x/L$ and mapped onto $r\in[0,1]$. Therefore, the ODE becomes
\begin{equation}
    (h^2+\lambda h)h''' =L^3\delta, \label{migODENum}
\end{equation}
and the volume condition becomes
\begin{equation}
    \int_0^1h\,\mathrm{d}r=\frac{2}{3L}.    \label{volConNum}
\end{equation}
To handle \eqref{volConNum}, the third-order ODE (\ref{migODENum}) is transformed into four first-order ODEs that write
\begin{equation}
    \begin{cases}
        f_1'=f_2,\\
        f_2'=f_3,\\
        f_3'=\frac{L^3\delta}{h^2+\lambda h},\\
        f_4'=f_1,
    \end{cases}\quad \text{where}\quad
    \begin{cases}
        f_1=h,\\
        f_2=h',\\
        f_3=h'',\\
        f_4= \int_{0}^{r}h\, \mathrm{d}r.
    \end{cases}
    \label{eq:2Dsys}
\end{equation}
The BCs (\ref{migBC}) cannot be directly implemented since the ODE for $f_3$ encounters singularity at the contact lines. Following \citet{benilov2015}, this difficulty is resolved by imposing a cut-off at the inner region $r=\Delta$ and $\ 1-\Delta$ with $\Delta\ll\lambda$, where new BCs are obtained by a Frobenius-style expansion as
\begin{align}
    h&\sim Lr-\frac{\delta}{2\lambda}L^2r^2\ln{Lr}+A_0L^2r^2+\mathcal{O}\left[r^3\ln{r}\right], \quad \text{as}\quad r\to0, \label{recBC}\\
    h&\sim KL(1-r)-\frac{L^2\delta}{2\lambda K}(1-r)^2\ln{[KL(1-r)]}\notag\\
        &+A_1K^2L^2(1-r)^2+\mathcal{O}\left[(1-r)^3\ln{(1-r)}\right], \quad \text{as}\quad r\to 1 \label{advanBC}.
\end{align}
These asymptotic relations provide six Dirichlet BCs for $f_1$-$f_3$ at $r=\Delta$ and $r=1-\Delta$. For $f_4$, we use the leading order of \eqref{recBC} and \eqref{advanBC} to obtain the BCs with an $\mathcal{O}(\Delta^3\ln\Delta)$ precision
\begin{equation}
    f_4(\Delta)=\frac{L\Delta^2}{2},\quad f_4(1-\Delta)=\frac{2}{3L}-\frac{KL\Delta^2}{2}.
\end{equation}
The eight BCs are sufficient to solve for $f_1$-$f_4$, and the unknown parameters $L$, $\delta$, $A_0$ and $A_1$. For perfect wetting, the asymptotic solution at the advancing contact line is replaced with
\begin{equation}
    h = \sqrt{\frac{8L^3\delta}{3\lambda}}(1-r)^{\frac{3}{2}}+A_1\lambda \left[\frac{L\delta^{1/3}(1-r)}{\lambda}\right]^{\frac{5+\sqrt{13}}{4}}, \quad \text{as}\quad r\to 1.        \label{advanBCper}
\end{equation}
The first-order system \eqref{eq:2Dsys} together with the BCs is solved using the MATLAB routine bvp5c.

\section{Numerics for 2D asymmetric spreading}\label{appD}
The numerical solutions of the PDE (\ref{2Dpde}) and BCs (\ref{sprBC}) are obtained by employing a similar numerical approach presented in \cite{qin2024} and \cite{Wangxl2025}, with an additional treatment of the pinned contact line. The advancing contact line at $x=L(t)$ introduces a moving boundary, which is addressed by introducing $r=x/L$. The governing equation on the rescaled domain $r\in[0,1]$ becomes
\begin{equation}
    \partial_th-\frac{r}{L}\dot L\partial_r h +\frac{1}{L^4}\partial_r\left[h^2(h+\lambda)\partial_r^3h\right]=0. 
\end{equation}
This equation can be transformed into a first-order system
\begin{equation}
    \begin{cases}
        f_1'=f_2,\\
        f_2'=f_3,\\
        f_3'=f_4,\\
        f_4'=\frac{-L^4\partial_th+L^3\dot Lrf_2-(3f_1+2\lambda )f_1f_2f_4}{f_1^2(f_1+\lambda)},
    \end{cases}\quad \text{with}\quad
    \begin{cases}
        f_1=h,\\
        f_2=h',\\
        f_3=h'',\\
        f_4=h'''.
    \end{cases}
\end{equation}
Here a prime denotes a derivative with respect to $r$. The cut-off in Appendix \ref{appC} is also implemented to avoid singularities in the last equation for both moving and pinned contact lines. The asymptotic relations at the advancing contact line is (\ref{advanBC}) for  finite $K$, or (\ref{advanBCper}) for $K=0$. Following \citet{eggers2025}, $h$ near the pinned contact line can be expanded as 
\begin{equation}
    h\sim\theta Lr+aL^2r^2-\frac{\dot{\theta}}{12\lambda\theta^2}L^3r^3,\quad \text{as}\quad r\to0.  
    \label{hp}
\end{equation}

The BCs of $f_1$ to $f_4$ can be evaluated using these asymptotic expressions. The time derivative of $h$, $L$, and $\theta$ is discreted with the first-order Euler implicit scheme. Given the solutions at the pervious time step, this ODE system of $f_1$-$f_4$ with four additional unknown parameters $L$, $\theta$, $A_1$ and $a$ can be solved with eight BCs of $f_1$-$f_4$ at $r=\Delta$ and $r=1-\Delta$.

\end{appen}

\bibliographystyle{jfm}
\bibliography{jfm}


\end{document}